%% file: TokyoTech_minor_revision_v1.tex
\documentclass[lettersize,journal]{IEEEtran}
\IEEEoverridecommandlockouts 

\usepackage{verbatim}
\usepackage{hyperref} 
\usepackage{graphicx, caption}
\usepackage{subcaption}
\usepackage{amssymb}
\usepackage{amsmath}
\usepackage{amsthm}
\usepackage{mathtools}
\usepackage{cite}
\usepackage{stfloats}
\usepackage{epstopdf}
\usepackage{psfrag}
\usepackage[mathscr]{euscript}
\usepackage{acronym}  
\usepackage{booktabs} 
\usepackage[linesnumbered,ruled,vlined]{algorithm2e}
\usepackage[table]{xcolor}
\usepackage{booktabs}

\SetKwInput{KwInput}{Input}
\SetKwInput{KwOutput}{Output}

\usepackage{color}
\usepackage{dsfont}
\usepackage{bbm}

\input{Support/acronym}
\input{Support/defmetric_thesis}

\title{Transformer Architecture with Minimal Inference Latency for Multi-Modal Wireless Networks}
\author{  		\IEEEauthorblockN{
			Minsu~Kim, Walid~Saad \textit{Fellow, IEEE}, Kui~Wang, Zongdian~Li \textit{Member, IEEE}, Tao~Yu \textit{Member, IEEE}, and Kei~Sakaguchi \textit{Senior Member, IEEE}
		}
        
        		\thanks{       			
        			M.\ Kim is with MediaTek Inc. USA, Warren, NJ, USA (email: minsu.kim@mediatek.com). \\
        			W. \ Saad is with the Institute for Advanced Computing,  Virginia Tech, Alexandria, VA, USA (email: walids@vt.edu. \\
        			Z. \ Li is with the State Key Laboratory of Blockchain and Data Security, Zhejiang University, Hangzhou, Zhejiang 310027, China (e-mail: zongdianli@zju.edu.cn)
        			K.\ Wang, T.\ Yu, and K.\ Sakaguchi are with Department of Electrical and
        			Electronic Engineering, Institute of Science Tokyo, Tokyo 152-8550, Japan (email: \{\texttt{kuiw, yutao, sakaguchi}\})@mobile.ee.titech.ac.jp )
					
					This research was supported by the U.S. National Science Foundation under Grant CNS-2210254.
					
				The implementation code is available on https://github.com/news-vt.
		
				}
        
        }

\begin{document}
\maketitle

\begin{abstract}
Next-generation wireless networks are expected to leverage multi-modal data sources in order to execute various wireless communication tasks such as beamforming and blockage prediction with situational-awareness. To do so, multi-modal transformers emerged as an effective tool,  however, existing transformer-based approaches suffer from high inference latency and large memory footprints when processing multi-modal data. Hence, such existing solutions cannot handle wireless communication tasks that require fast inference to track a dynamically changing environment with moving vehicles and blockages. One major bottleneck is the reliance on attention mechanisms whose complexity grows quadratically with respect to the number of tokens. Hence, in this paper, a novel, fast multi-modal transformer inference framework is designed to practically support wireless communication tasks by processing only important tokens. To this end, an optimization problem is formulated to find the optimal number of tokens under a target floating point operations (FLOPs) for a given wireless communication task while maintaining the task accuracy. To solve this problem, modality-specific tokenizers are first designed to project each modality into the same embedding dimension. Then, a token router is introduced to learn the importance of each token and process only important tokens. Subsequently, a trainable keep ratio is introduced to learn how many tokens should be processed for each layer under the target FLOPs. Simulation results show that, on DeepSense 6G beamforming tasks, the proposed framework can reduce the inference latency, GPU memory, and FLOPs by 86.2\% 35\%, and 80\%, respectively, with negligible accuracy loss compared to a baseline that processes all tokens. To further validate the feasibility of the proposed framework for real-world deployments, a multi-modal handover dataset is developed using a real-world testbed. Emulation results on the developed dataset show that the proposed framework can proactively initiate handover before blockage, while the baseline experiences significant received signal strength degradation due to higher inference latency. 

\end{abstract}

\section{Introduction}
Next-generation wireless networks such as 6G can potentially leverage multiple sensing modalities to provide seamless connectivity and higher throughput in dynamic wireless environments via situational-awareness \cite{10929033}. This, in turn, led to the proliferation of transformer-based methods for executing a variety of wireless communication tasks \cite{alkhateeb2023deepsense}. For instance, in vehicular networks, multi-modal transformers have been actively used to aid high frequency beam selection and blockage prediction by fusing multiple sensing modalities (e.g., LiDAR, camera, radar) \cite{park2025resource}. However, multi-modal transformers often suffer from significant inference latency and memory usage, which limit their applicability in real-world wireless communication tasks. One major inefficiency bottleneck in transformer architecture is the reliance on attention mechanisms whose complexity grows quadratically with respect to the number of tokens. Although multi-modal data can provide more information about the current environment, it also increases the number of tokens that must be processed by transformers. As a result, the predictions of transformers can fail to satisfy real-time requirements due to high inference latency. For instance, consider a proactive \ac{mmWave} handover task based on multi-modal data as considered in \cite{10735366} and \cite{11018220}. Although a trained transformer can accurately predict an upcoming obstacle, its predictions will be obsolete if it finishes inference after the obstacle blocks \ac{LoS} paths.  One can address this challenge by simply reducing the number of tokens in transformer blocks. However,  this approach ignores the importance of each token and inter-modal correlation across tokens from different modalities, thereby failing to predict upcoming obstacles. Therefore, there is a need to design an efficient multi-modal transformer inference framework \emph{that can balance performance and latency} by finding the optimal number of tokens to satisfy real-time requirements of wireless communication tasks. 

There has been a number of recent works \cite{rao2021dynamicvit, bolya2023token, bonnaerens2023learned, wu2024videollm, you2025layer} that focused on reducing the complexity of attention mechanisms. The work in \cite{rao2021dynamicvit} proposed token pruning by removing redundant tokens by estimating an importance score. In \cite{bolya2023token}, the authors proposed token merging by gradually combining similar tokens from images. The authors in \cite{bonnaerens2023learned} combined token pruning and merging for vision tasks. The work in \cite{mod} proposed \ac{MoD}, where token-level routing is introduced to process only Top-K important tokens while bypassing other tokens for language tasks. In \cite{wu2024videollm}, the authors used \ac{MoD} in large language models (LLMs) to reduce the number of vision tokens in video tasks. The authors in \cite{you2025layer} applied \ac{MoD} to diffusion transformers with differentiable token compression ratios. However, the prior works in \cite{rao2021dynamicvit, bolya2023token, bonnaerens2023learned, wu2024videollm, you2025layer} were proposed to remove tokens from a single modality, e.g., vision or language. Hence, their approach cannot be directly applicable to multi-modal wireless communications scenarios. Moreover, the works in \cite{rao2021dynamicvit, bolya2023token, bonnaerens2023learned, wu2024videollm} controlled the number of tokens heuristically based on prior observations instead of optimizing the number of tokens and the performance together. Meanwhile, the work in \cite{10171192} proposed a framework that can find optimal quantized neural networks without jeopardizing accuracy.  The authors in \cite{kim2024spafl} proposed a framework that can find optimal structured sparsity for transformers. Although reducing the complexity of attention modules can be done via quantization \cite{10171192} and pruning \cite{kim2024spafl}, these approaches still experience quadratically increasing complexity for long sequence of tokens from multi-modal data. 

Multi-modal transformers have been adopted in wireless communications tasks such as beamforming \cite{cui2024sensing, park2025resource, 10735366, 11018220} and blockage prediction \cite{10660494, 10680020} to leverage a broad range of sensing modalities. In \cite{cui2024sensing}, the authors proposed a transformer-based \ac{mmWave} beamforming based on multi-modal data including image, LiDAR, radar, and GPS. The work in \cite{park2025resource} employed knowledge distillation to train a student model with a limited modality by leveraging a teacher model trained with multiple modalities. The authors in \cite{10735366} proposed a two-step beamforming method in which a multi-modal transformer first predicts a group of optimal beams and then uses reinforcement learning to decide the optimal beam. In \cite{11018220}, the authors investigated a training method to achieve strong generalization to weather changes for multi-modal transformer with beamforming tasks. In \cite{10660494} and \cite{10680020}, the authors investigated the use of multi-modal transformers with vision and GPS to predict blockages in mmWave vehicular scenarios. However, the prior works in \cite{11018220,  park2025resource}, and \cite{cui2024sensing, 10660494, 10680020} did not consider computing efficiency (e.g., inference latency) of their transformer models and their feasibility for real-world deployments. Moreover, the authors in \cite{10660494} exhibit quadratically increasing inference latency with respect to the input window size, making it challenging to promptly predict blockages. Although in \cite{10735366}, the authors discussed the efficiency of their model, reporting the inference latency, they did not develop any approach to improve efficiency and reduce inference time. To the best of our knowledge, there are no current works that jointly optimized wireless network performance and the number of transformer tokens under a target latency in multi-modal wireless communication scenarios.

The main contribution of this paper is a novel multi-modal transformer inference framework that can optimize the number of tokens and model together under a target computational budget for deployment in practical wireless communication use cases. We first define a class of multi-modality-aided beam and link status prediction problems in a multi-\ac{V2I} environment with a transformer-based model. Although existing transformer-based approaches can achieve high accuracy for these problems, transformer models suffer from high inference latency. 
One major bottleneck of transformer-based approach is the attention mechanism whose complexity grows quadratically with respect to the number of tokens from multi-modal input. Hence, we formulate an optimization problem whose goal is to control the number of tokens so as to maximize wireless network performance under a \ac{FLOPs} target. Here, a target \ac{FLOPs} captures a computational budget that can support a task specific latency requirement, such as beam coherence time.  We first project all modalities into the same embedding dimension by designing tokenizers. Then, we introduce a token router to learn the importance of each token from different modalities in every transformer block. We process Top-$\topk$ important tokens while other tokens bypass the entire transformer block, thereby reducing the complexity of transformer blocks as $\mathcal{O}(\topk^2)$. To optimize $\topk$, we define a learnable keep ratio, which represents the rate of tokens that need to be processed by each transformer block. We map every keep ratio to its two nearest integers for the Top-$\topk$ operation and perform two forward passes with the mapped integers. To generate gradient paths to keep ratios, we linearly combine the outputs of the two forward passes. Hence, during training, we optimize keep ratios, token routers, and model parameters together in an end-to-end manner. Simulation results demonstrate the performance and efficiency of our framework compared to a baseline that processes every token on two multi-modal mmWave beamforming datasets. For instance, our framework can improve the inference latency, GPU memory, and \ac{FLOPs} by 86.2\% 35\%, and 80\%, with only negligible accuracy loss on DeepSense 6G beamforming tasks compared to the baseline. We further validate our approach on multi-modal handover datasets collected in a real-world testbed at the Science Tokyo campus. Emulation results over this real dataset show that our framework can perform proactive handover before blockages without experiencing \ac{RSSI} damage. Meanwhile, the \ac{RSSI} of the baseline significantly degrades by 70\% due to blockages and obsolete predictions.

The rest of this paper is organized as follows. Section \ref{sec: tokyo_system} presents the system model. Section \ref{sec: proposed} presents our framework.  In Section \ref{sec: tokyo_experiments}, simulation results are provided. Finally, conclusions are drawn in Section \ref{sec: tokyo_conclusion}. 
Notation: An overview of our notation is shown in Table \ref{tab: notation}.

\begin{figure}[t]
	\centering 				%
	\begin{subfigure}[t]{0.49\linewidth}
		\centering
		\includegraphics[width=\textwidth]{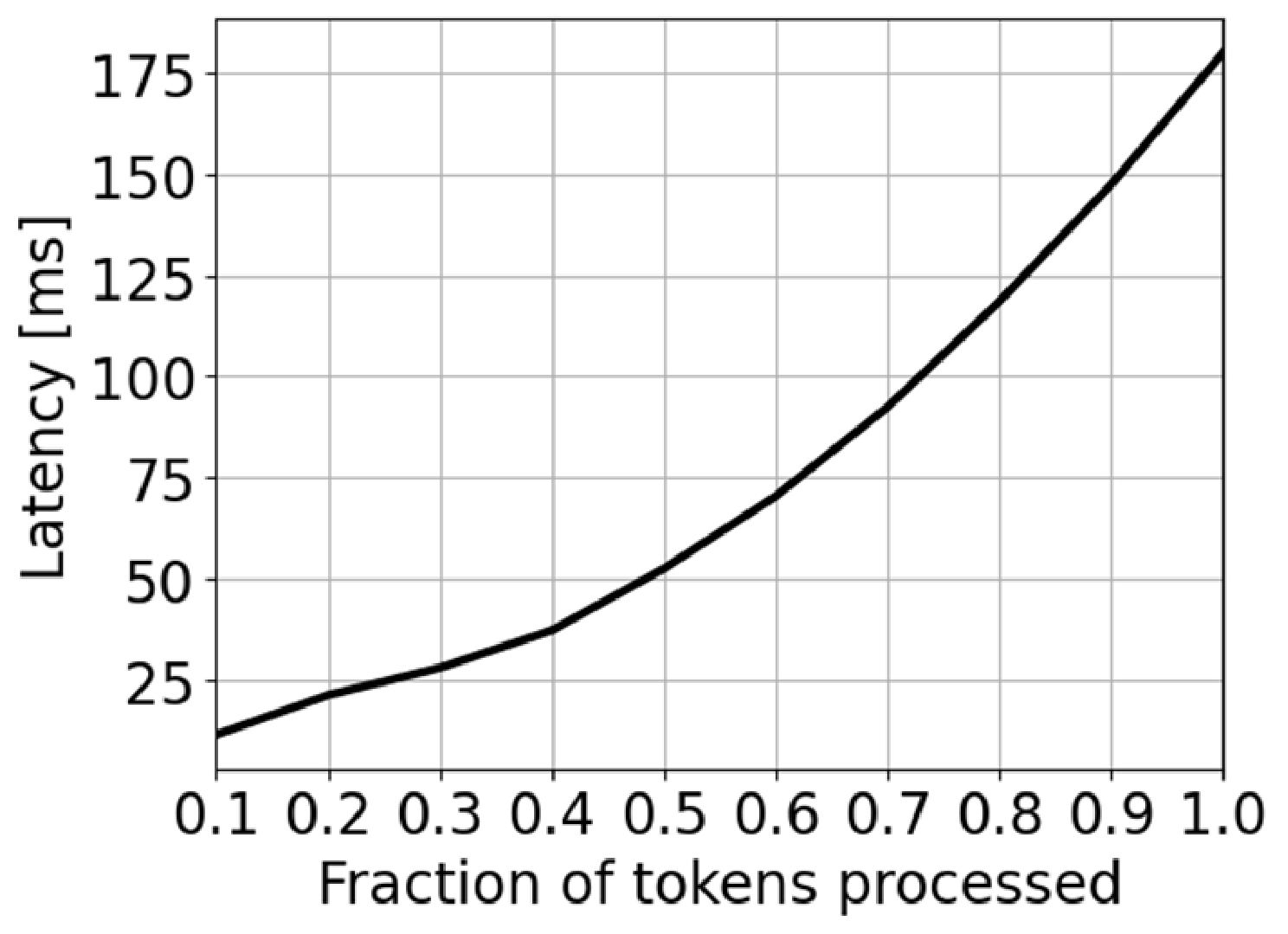}
		\caption{Latency for increasing the fraction of processed tokens.}
	\end{subfigure}
	\begin{subfigure}[t]{0.49\linewidth}
		\centering
		\includegraphics[width=\textwidth]{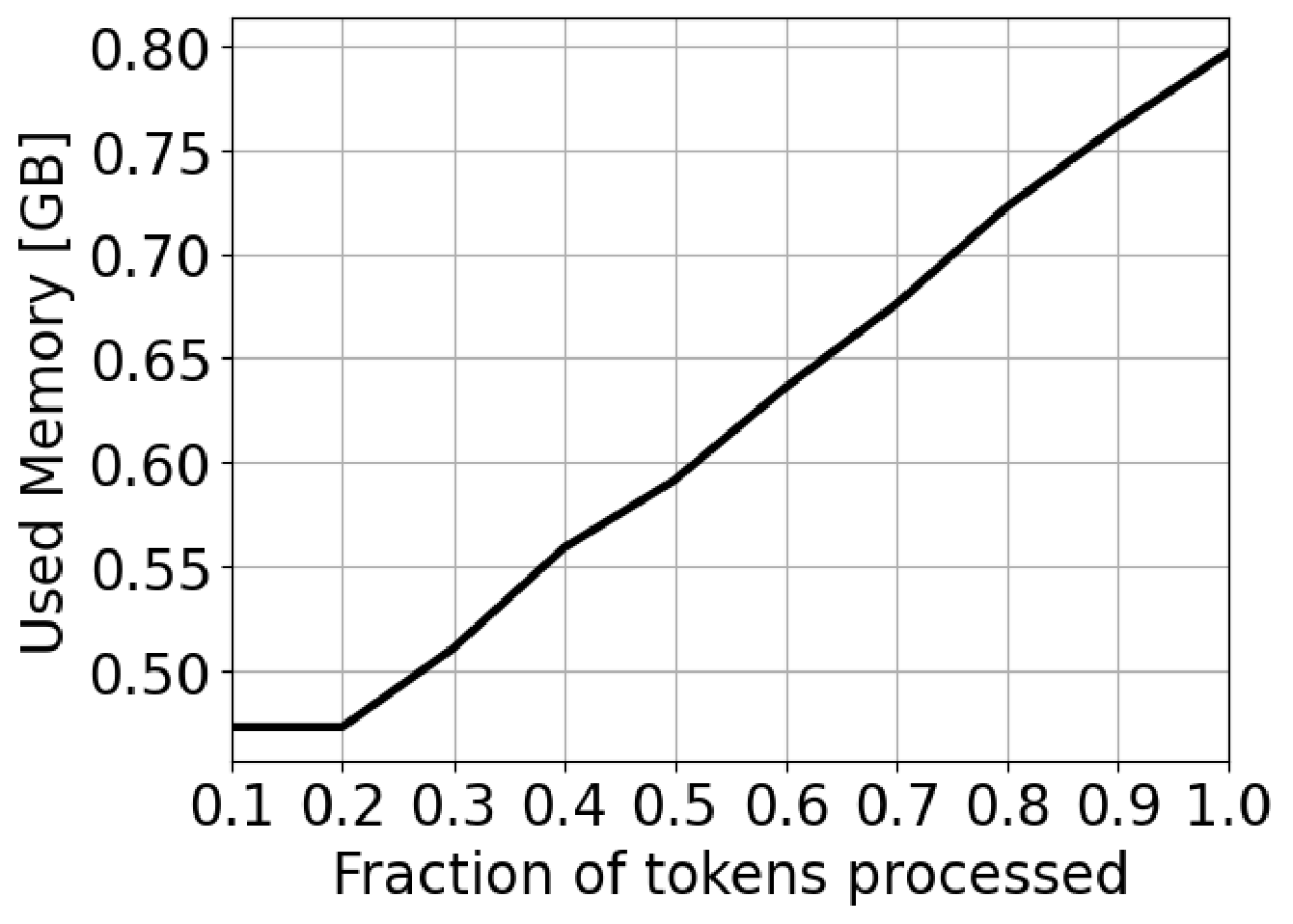}
		\caption{GPU memory usage for increasing the fraction of processed tokens.}
	\end{subfigure}
	
	\caption{Latency and GPU memory usage for increasing the number of processed tokens on an A30 GPU.}
	\label{fig: intro_latency_memory}
\end{figure}

\begin{table}
	\caption{Summary of our key notations.} \label{tab: notation}
	\begin{center}
		\renewcommand{\arraystretch}{1.3}
		\begin{tabular}{c p{6cm} }
			\hline 
			{\bf Notation} & {\hspace{2.5cm}}{\bf Definition}
			\\
			\midrule
			\hline
			$\nn$ & Transformer-based ML model  \\ \addlinespace			\hline 
			$\tokentotalnum$ & Total number of tokens \\ \addlinespace			\hline 
			$\topk$ & Number of selected tokens \\ \addlinespace			\hline 
			$\tokyoloss $ & Loss function\\ \addlinespace			\hline 
			$\flopbound$ & Maximum computational bound\\ \addlinespace 			\hline 
			$\bar{\rawmodalin}$ & Raw multi-modal data input sequence\\ \addlinespace			\hline 
			$\multimodalin$ 	& Multi-modal input token sequence \\ \addlinespace			\hline 
			$\score_l$		& Token importance scores in transformer block $l$ \\ \addlinespace			\hline 
			$\keepratio$ 		& Keep ratio	\\ \addlinespace			\hline 
			$\flopbound'$ & Target keep ratio \\ \addlinespace			\hline 
			$\beam$ & Beamforming vector \\ \addlinespace			\hline 
			$\eta$ & Path loss exponent \\ \addlinespace			\hline 
			$J_v$ & Number of signal propagation paths from the RSU to vehicle $v$ \\ \addlinespace			\hline 
			$\xi$ 	& Large scale fading \\ \addlinespace			\hline 
			$\tokyolayernum$	& Number of transformer blocks \\ \addlinespace			\hline 
			$\tokyolink$ & Link status \\ \addlinespace		
			\hline 
			
		\end{tabular}
	\end{center}\vspace{-0.63cm}
\end{table}%

\section{System Model} \label{sec: tokyo_system}
We consider a \ac{V2I} environment that includes a \ac{RSU} with $\tokyonumatennta$ \ac{mmWave} uniform planar array (UPA) antennas and a set $\mathcal{V}$ of $V$ vehicles with a single \ac{mmWave} antenna. The \ac{RSU} employs a predefined beamforming codebook $\codebook = \{\beam_i\}_{i=1}^{|\codebook|}$, where $\beam_i \in \mathbb{C}^\symboldim$ is a beamforming vector.
Then, the total \ac{RSS} of vehicle $v$ with beamforming vector $\beam_i$ will given by: \cite{park2025resource, 9838838}
\begin{align}
	\tokyorss_v(\beam_i) = \sum_{j=1}^{J} \big( \tokyotxpower_{v}^j(\beam_i) - \Omega_{v}^j \big),
\end{align}
where $J_v$ is the number of signal propagation paths between the \ac{RSU} and vehicle $v$, $\tokyotxpower_{v}^j(\beam_i)$ is the magnitude of the array response over path $j$, and $ \Omega_{v}^j$ is the corresponding path loss. Here,  $\tokyotxpower_{v}^j(\beam_i)$ can be given as
\begin{align}
	\tokyotxpower_{v}^j(\beam_i) = 10\log_{10} |\beam_i \boldsymbol{a}_v(\theta_j, \phi_j)|,
\end{align}
where $\boldsymbol{a}_v(\theta_j, \phi_j)$ is the array response vector with azimuth angle $\theta_j$ and elevation angle $\phi_j$
\begin{align}
	\boldsymbol{a}_v(\theta_j, \phi_j) = \boldsymbol{a}_{v, x} (\theta_j, \phi_j) \otimes  \boldsymbol{a}_{v, y}(\theta_j, \phi_j).  
\end{align}
For a half-wavelength-spacing, the array response vector on each axis will be:
\begin{align}
	&\boldsymbol{a}_{v, x} (\theta_j, \phi_j) = \frac{1}{\sqrt{\tokyonumatennta}} [1, e^{j \sin(\theta_j) \cos(\phi_j)}, ..., e^{j  (\tokyonumatennta - 1) \sin(\theta_j) \cos(\phi_j)}], \ka
	&\boldsymbol{a}_{v, y} (\theta_j, \phi_j) = \frac{1}{\sqrt{\tokyonumatennta}} [1, e^{j \sin(\theta_j) \sin(\phi_j)}, ..., e^{j  (\tokyonumatennta - 1) \sin(\theta_j) \sin(\phi_j)}].
\end{align}
The path loss will be given by:
\begin{align}
	\Omega_{v}^j = \tokyopower_0 + 10\eta\log_{10}(\tokyodistance_v^j) + \xi,
\end{align} 
where $\tokyopower_0$ is a reference path loss at 1 m, $\eta$ is the path loss exponent, $\tokyodistance_v^j$ is the distance of path $j$, and $\xi$ is the large-scale fading coefficient \cite{10177877}. 

We assume that the \ac{RSU} has multi-modal sensors, including camera, LiDAR, radar, and GPS. To capture cross-modal correlations of multi-modal input $\rawmodalin$, the \ac{RSU} deploys a transformer-based \ac{ML} model $\nn$. We assume that model $\nn$ has $\tokyolayernum$ transformer blocks that have self-attention, MLP, residual connections, and layer normalization. At every sampling interval $t \in \mathcal{T}$, the \ac{RSU} makes a network-level decision (e.g., beamforming or handover) based on $\tokyoframe$ past multi-modal samples $\bar{\rawmodalin} [t] = [\rawmodalin[t - \tokyoframe+1], ... , \rawmodalin[t]]$ as done in \cite{alkhateeb2023deepsense, yang2023environment}. For a beamforming task, at every $t$, the goal of $\nn$ is to find the next optimal beamforming vector $\beam^*[t+1]$ that can maximize the sum of \ac{RSS} over vehicles for given input $\rawmodalin[t]$ as follows \cite{cui2024sensing, park2025resource}
\begin{align}
	&\max_{\nn} \quad \P \big [\nn(\bar{\rawmodalin}[t]) = \beam^*[t+1] \big] \\
	& \ \text{s.t.} \quad \  \beam^*[t+1] = \arg\max_{\beam} \sum_{v=1}^{V} \tokyorss_v(\beam[t+1]). \label{prob: beamforming}
\end{align}
For a handover task, $\nn$ predicts the next link status of each vehicle at every $t$ for given input $\bar{\rawmodalin}[t]$ as \cite{11018220}
\begin{align}
	\tokyolink_v[t+1] = \begin{cases}
		1, & \text{if} \ \tokyorss_v[t+1] > \tokyorss_{\text{th}}, \\
		0, & \text{otherwise},
	\end{cases}
\end{align}
where $\tokyorss_{\text{th}}$ is a threshold \ac{RSS} to initiate handover. Then, the goal of $\nn$ will be maximizing the probability of accurately predicting the link status of each vehicle, i.e., we need to solve:
\begin{align}
	\max_{\nn} \quad \prod_{v=1}^{v} \P \big [\nn(\bar{\rawmodalin}[t]) = \tokyolink_v[t+1] \big]. \label{prob: handover}
\end{align}
Although transformer model $\nn$ can solve problems \eqref{prob: beamforming} and \eqref{prob: handover} as shown in \cite{cui2024sensing, park2025resource, 10735366, 11018220}, there is no guarantee that inference can be executed within each time frame. Multi-modal data with multiple past frames increase the number of tokens that need to be processed by $\nn$. Specifically, for an input token sequence $\multimodalin \in \mathbb{R}^{\tokentotalnum\times \tokyodimension}$ with $\tokentotalnum$ tokens from $\bar{\rawmodalin}[t]$ and embedding dimension $d$, the complexity of $\nn$ increases as $\mathcal{O}(\tokentotalnum^2)$. One major bottleneck is the self-attention mechanism whose complexity  grows quadratically with respect to $\tokentotalnum$ \cite{bolya2023token} as shown in Fig. \ref{fig: intro_latency_memory}. Hence, after $\nn$ makes predictions, the optimal beam or link status could be already changed due to high inference latency. Therefore, it is important to process only important tokens to minimize the inference latency while maintaining the performance of $\nn$.   

\subsection{Problem Formulation} \label{subsec: prob}
We now formulate our optimization problem to minimize a task-specific loss function by optimizing the number of tokens $\boldsymbol{\topk}= [\topk_1, \topk_2, ..., \topk_\tokyolayernum]$ for $\tokyolayernum$ transformer blocks under a target computational budget. Here, the loss can be a cross-entropy and a binary cross-entropy loss for beamforming and handover tasks, respectively. A tradeoff exists between the performance of $\nn$ and the inference latency with respect to $\boldsymbol{\topk}$. A very large $\boldsymbol{\topk}$ increases the inference latency, resulting in obsolete predictions for dynamically changing vehicular environments. Meanwhile, a small $\boldsymbol{\topk}$ can decrease the performance of $\nn$ by failing to accurately predict beamforming vectors $\beam[t+1]$ or link status $\tokyolink[t+1]$. Therefore, it is important to optimize $\nn$ and $\boldsymbol{\topk}$ under a target computational budget to practically support dynamic wireless environments. This can be achieved by solving the following problem:
%
%
\begin{align}
	&\min_{\nn, \boldsymbol{\topk}} \ \tokyoloss(\nn, \boldsymbol{\topk}) \label{problem} \\
	& \ \text{s.t.} \ \text{FLOPs}(\nn, \boldsymbol{\topk}) \leq \flopbound \label{constraint_1} \\
	& \quad \ \ \topk_l \leq \tokentotalnum, \forall l \in [1, ..., \tokyolayernum], \label{constraint_2}
\end{align}
where $\tokyoloss(\nn, \boldsymbol{\topk})$ is a task specific loss function, $\text{FLOPs}(\cdot)$ measures the number of \ac{FLOPs} for given $\nn$ and $\boldsymbol{\topk}$, and $\flopbound$ is the maximum computational bound to support the current wireless communication task. Here, we use \ac{FLOPs} as a surrogate for inference latency  \cite{kwon2022fast, zheng2022savit, park2023accurate, meng2024falcon} in our optimization problem because real inference latency heavily depends on hardware, processor (e.g., CPU or GPU) utilization, memory access, and implementation code structure. Hence, real inference time cannot be expressed as a closed form due to the analytically intractable factors. Constraint \eqref{constraint_1} captures the target latency in terms of the number of \ac{FLOPs}.  For instance, it can be the maximum \ac{FLOPs} needed for complete inference within a beam coherence time \cite{va2016impact}\footnote{For a MISO link with a vehicle moving at 20 meter per second with $\tokyodistance = 2.5$ m, the beam coherence time is around 100 ms \cite{khorsandmanesh2024beam}}.  Constraint \eqref{constraint_2} represents the maximum number of tokens that can be processed by each transformer block. 

Problem \eqref{problem} is NP-hard due to $\nn$, $\boldsymbol{\topk}$, and the non-convexity of the loss function. It is also challenging to analytically model the relationship between $\boldsymbol{\topk}$ and $\tokyoloss(\nn, \boldsymbol{\topk})$ because $\nn$ and $\boldsymbol{\topk}$ are correlated. Moreover, it is difficult to measure the importance of each token with different modalities because we need to consider both the inter-modal correlation and per-modal specificity. Next, we present the proposed transformer architecture approach for solving problem \eqref{problem}.

\section{Transformer Design For Minimal Latency} \label{sec: proposed}
We first provide an overview of the proposed framework to solve the formulated problem \eqref{problem}. Then, we provide its three main components: 1) Projecting all modalities onto the shared embedding dimension, 2) Measuring the importance of each token based on a token router, and 3) Differentiable keep ratios to optimize the number of tokens $\topk_l$ to be processed in each transformer block $l, \forall l$. 

\subsection{Overview of the Proposed Framework}
\begin{figure*}[t] 		
	\centering 				%
	\begin{subfigure}[t]{0.49\linewidth}
		\includegraphics[width=\textwidth]{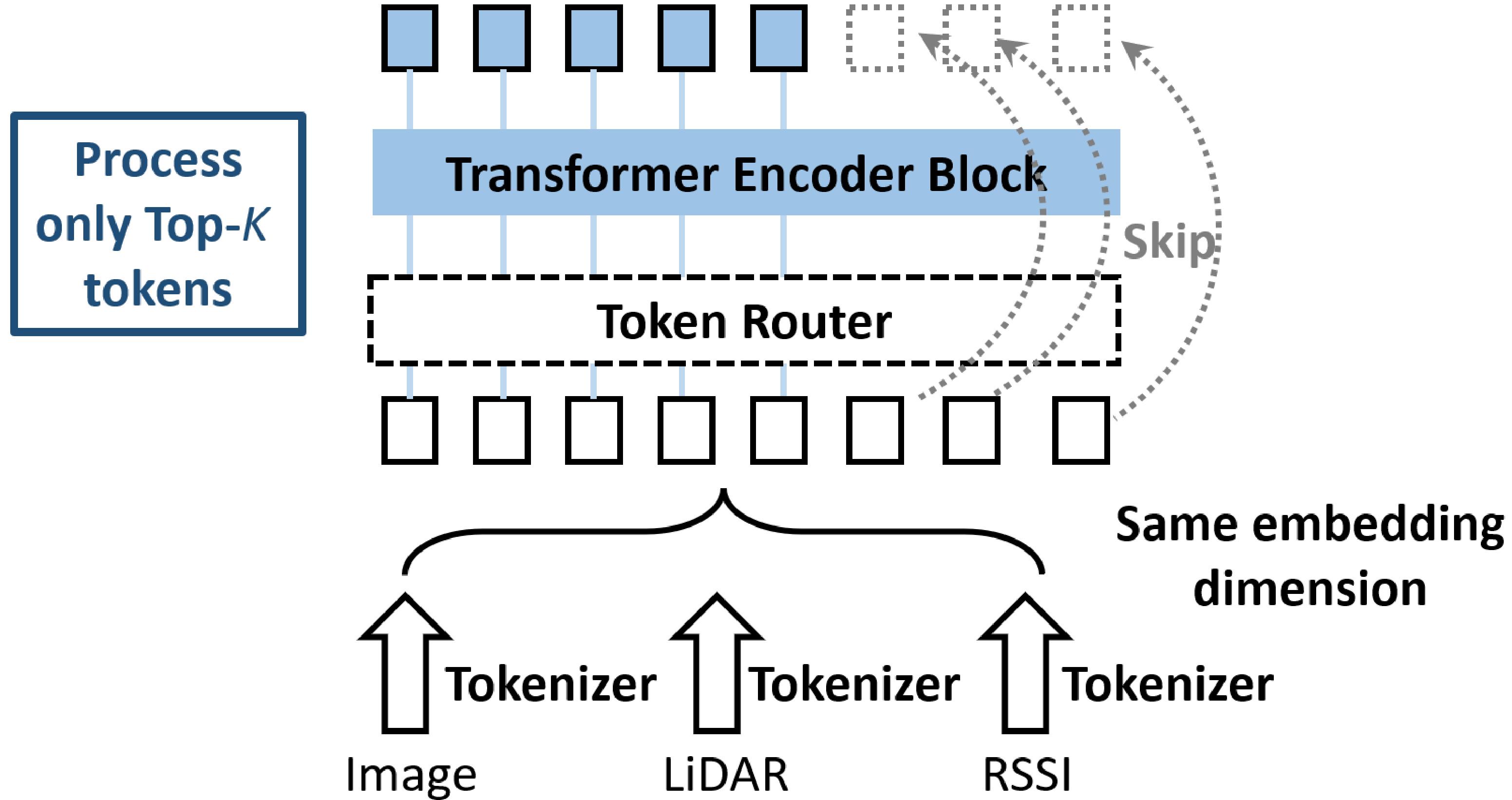}
		\caption{Projecting all modality into the same embedding  dimension and predicting the importance of each token using the token router.}
		\label{fig: token_router}
	\end{subfigure}
	\hfill
	\begin{subfigure}[t]{0.49\linewidth}
		\includegraphics[width=\textwidth]{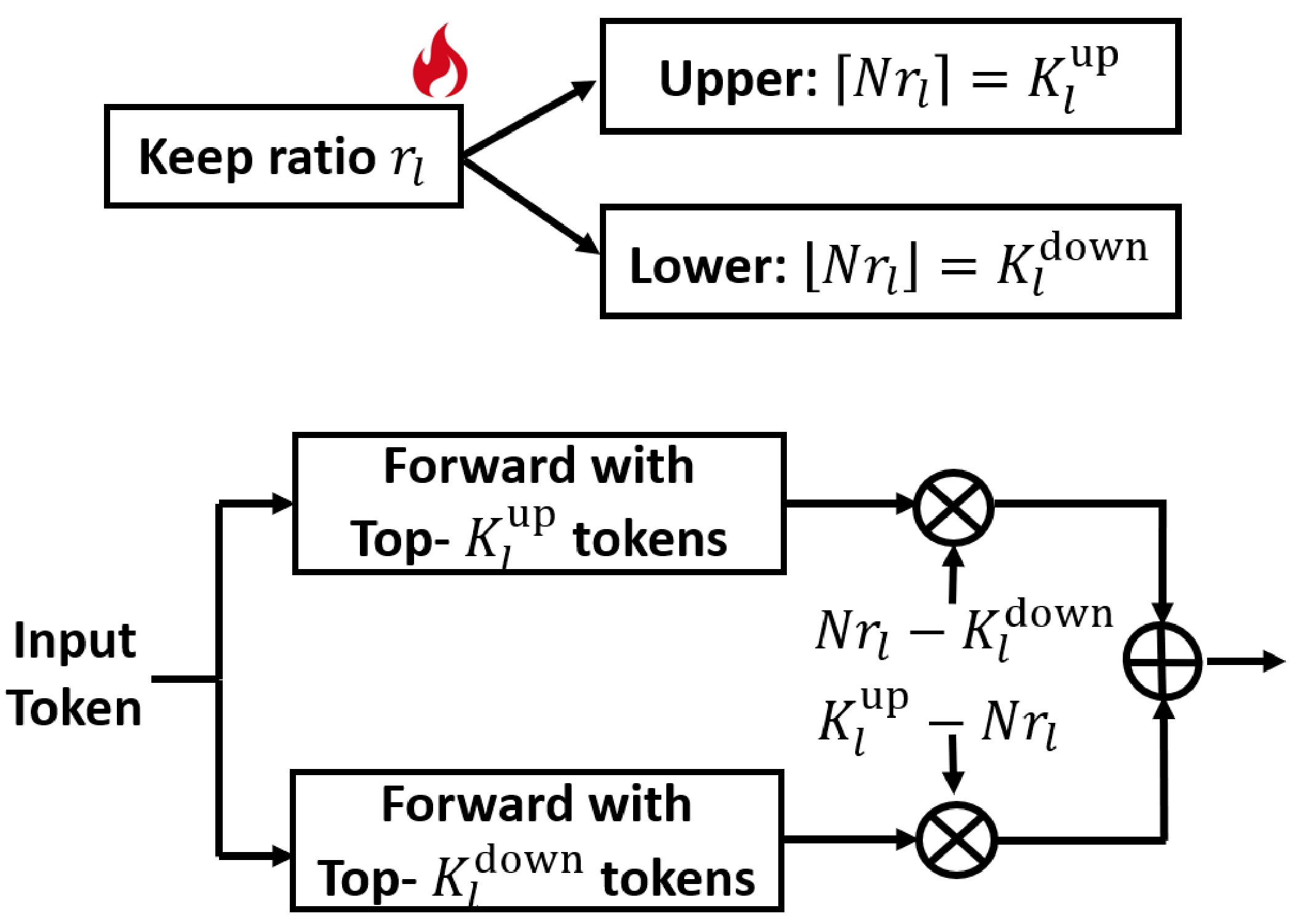}
		\caption{Trainable keep ratio parameters to optimize $\boldsymbol{\topk}$.}
		\label{fig: keep_ratio}
	\end{subfigure}
	\caption{Overview of the proposed framework.}
	\label{fig: proposed}
\end{figure*}

Fig. \ref{fig: proposed} provides an overview of the proposed framework. First, our framework utilizes light-weight tokenizers that map multi-modal inputs into the shared embedding dimension. Hence, each modality input is transformed into token sequences that are in the same space as other modalities, thereby decreasing the inherent modality gap. Then, all tokens are concatenated and fed into shared transformer blocks to extract representations in the same embedding dimension. To learn the importance of each token, each transformer block uses a light-weight token router that predicts the importance based on the embedding. Then, we only process Top-$\topk_l$ tokens in transformer block $l$ while bypassing computations for other tokens entirely for this block. Hence, we reduce the complexity from $\mathcal{O}(N^2)$ to $\mathcal{O}(\topk_l^2)$. To optimize $\topk_l, \forall l$, we define a trainable and continuous \emph{keep ratio} parameter, which represents the rate of tokens that need to be processed by each transformer block. During training, the keep ratio parameter is mapped to two nearest integer values $\topk_l^\text{up}$ and $\topk_l^\text{down}$ to perform two forward passes with Top-$\topk_l^\text{up}$ and Top-$\topk_l^\text{down}$ tokens. We combine the outputs of these two forward passes with the normalized distances between the keep ratio parameter and $\topk_l^\text{up}$ and $\topk_l^\text{down}$. This linear combination generates gradient paths to the keep ratio parameter. To satisfy constraint \eqref{constraint_1}, we use a \ac{MSE} loss between the keep ratio parameters and the target keep ratio to meet the computational requirement.

\subsection{Modality Tokenization} \label{subsec: tokenzier}
We use light-weight tokenizers to transform various modalities into token sequences in the same embedding dimension to compare the importance of each token. We consider images, point clouds, radar, GPS, and received signal strength index (RSSI) as examples because those modalities are widely used for multi-modality-assisted wireless communication systems \cite{9923616, yang2023environment, 10949588}.
\begin{itemize}
	\item \emph{Images}: Consider a 2D image input $\rawmodalin_{\text{img}} \in \mathbb{R}^{H\times W \times C}$, To tokenize $\rawmodalin_{\text{img}}$, we utilize a tiny CNN model that consists of a few ResNet blocks and one 1D-convolutional layer. The ResNet blocks extract semantic features of image inputs, and the 1D-convolutional layer projects the features into the embedding dimension as follows
	\begin{align}
		\rawmodalin_{\text{img}} \in \mathbb{R}^{H\times W \times C} \rightarrow \multimodalin_{\text{img}} \in \mathbb{R}^{\tokentotalnum_{\text{img}} \times \tokyodimension},
	\end{align}
	where $\tokentotalnum_{\text{img}}$ is the number of tokens from $\rawmodalin_{\text{img}}$,  and $\tokyodimension$ is the embedding dimension. 
	
	\item \emph{Point Cloud}: Consider point cloud data $\rawmodalin_{\text{pc}} = (\boldsymbol{x}_\text{pc}^{\text{p}}, \boldsymbol{x}_\text{pc}^{\text{f}})$, where $\boldsymbol{x}_\text{pc}^{\text{p}} \in \mathbb{R}^3$ is the 3D coordinates, and $\boldsymbol{x}_\text{pc}^{\text{f}}$ is the features of each point. We use a bird's eye view transformation to provide the locations of surrounding objects while removing ground reflections. Hence, we project 3D point clouds onto a 2D plane and remove unnecessary information. We then use a similar tiny \ac{CNN} model that is used for image inputs to transform point clouds into token sequences as follows
	\begin{align}
		\rawmodalin_{\text{pc}} = (\boldsymbol{x}_\text{pc}^{\text{p}}, \boldsymbol{x}_\text{pc}^{\text{f}}) \rightarrow \rawmodalin_{\text{pc}}' \in \mathbb{R}^{H \times W} \rightarrow \multimodalin_{\text{pc}} \in \mathbb{R}^{\tokentotalnum_{\text{pc}} \times \tokyodimension},
	\end{align}
	where $\tokentotalnum_{\text{pc}}$ is the number of tokens from $\rawmodalin_{\text{pc}}$.
	
	\item \emph{Radar}: We consider a frequency modulated continuous wave (FMCW) radar input frame $\rawmodalin_{\text{rad}}$ that consists of (number of receive antennas) x (number of samples per chirp) $\times$ (number of chirps per frame) \cite{alkhateeb2023deepsense}. We use the highest point sampling to process FMCW radar cubes into [velocity, azimuth, altitude, depth] and downsample it to $\tokentotalnum_{\text{rad}}$. We then utilize a few linear layers to project the $\tokentotalnum_{\text{rad}}$ samples into token sequences with  embedding dimension $\tokyodimension$ as follows 
	\begin{align}
		\rawmodalin_{\text{rad}} \rightarrow \rawmodalin_{\text{rad}}' \in \mathbb{R}^{\tokentotalnum_{\text{rad}} \times 4} \rightarrow \multimodalin_{\text{rad}} \in \mathbb{R}^{\tokentotalnum_{\text{rad}} \times \tokyodimension}.
	\end{align}
	
	\item \emph{GPS} and \ac{RSSI}: Since GPS and RSSI are scalar values, we use a few linear layers to tokenize those modalities. Hence, each GPS input $\rawmodalin_{\text{GPS}}$ and \ac{RSSI} input $\rawmodalin_{\text{RSSI}}$ is projected to one token of dimension $\tokyodimension$, respectively. 
\end{itemize}
Now, each modality input is transformed to token sequences that are in the same embedding dimension as other modalities, thereby decreasing the inherent modality gap. Then, for $\tokyomodalitynum$ modality inputs,  all tokens are concatenated as  $ \multimodalin = [\multimodalin_1, \multimodalin_2, ... ,\multimodalin_\tokyomodalitynum]$ and fed into shared transformer blocks to extract representations on the same space.

\subsection{Token-level Routing} \label{subsec: rotuer}
We now introduce our token-level routing scheme. 
First, we observe that not every token in $\multimodalin$ is important. For instance, image data $\rawmodalin_\text{img}$ usually has redundancy such as background information. Meanwhile, some tokens can have important relations with other modalities. To learn the importance of each token, we use \ac{MoD} \cite{mod} for token-level routing. Since we process tokens from multiple modalities, we remove the auxiliary loss and MLP predictor that were exclusively modeled for \ac{NLP} tasks from \ac{MoD}. To the best of the authors' knowledge, we are the first to apply the idea of token routing in \ac{MoD} to wireless domains with multi-modal data. 

To predict the importance of each token from different modalities, we deploy a light-weight token router in each transformer block as shown in Fig. \ref{fig: token_router}. Each token router consists of two linear layers with a GELU activation function.  For transformer block $l$, its token router predicts the importance of each token $\score_l$ as follows
\begin{align}
	\score_l = \text{Router}_l(\multimodalin) \in \mathbb{R}^\tokentotalnum.
\end{align}
Then, the token router chooses Top-$\topk_l$ tokens based on the predicted importance, and routes the selected tokens to the subsequent transformer block while other tokens bypass the transformer block entirely. Hence, the complexity of the transformer block reduces from $\mathcal{O}(\tokentotalnum^2)$ to $\mathcal{O}(\topk_l^2)$. To train the token router, we multiply the predicted importance of Top-$\topk_l$ tokens by the output activations of the transformer block $\transformer_l$ as follows
\begin{align}
	\bar{X_i} = \begin{cases}
		s_i \transformer_l(X_i) + X_i, & \text{if} \ s_i \in \text{Top-}\topk_l, \\
		X_i, & \text{else},  \label{router_grad}
		\end{cases}
\end{align}
where $X_i$ is input token $i \in [1, ..., \tokentotalnum]$. Hence, gradient paths to the token router can be generated during backpropagation by \eqref{router_grad}. Therefore, token routers can be optimized together with model $\nn$ during training. 

\subsection{Optimizing $\boldsymbol{\topk}$} \label{subsec: keep_ratio}
A tradeoff exists between the inference latency and the performance of model $\nn$ with respect to the number of tokens $\boldsymbol{\topk}$. We can improve the latency by processing a small number of tokens, but it often degrades the performance because it cannot extract enough representations from the tokens. One can heuristically optimize $\boldsymbol{\topk}$ by performing a number of hyperparameter optimizations or investigating which layer is important as done in \cite{mod, wu2024videollm}. However, this approach takes significant time due to the size of typical transformer models. Hence, motivated by \cite{you2025layer}, we define a trainable keep ratio parameter $\keepratio_l \in [0, 1]$ in transformer block $l, \forall l$. The keep ratio parameter $\keepratio_l$ decides how many tokens will be processed in transformer block $l$. We optimize $\keepratio_l$ together with token routers and $\nn$ during training, thereby finding optimal $\boldsymbol{\topk}$ for each layer in an end-to-end manner. 

We present our overall training pipeline for optimizing $\keepratio_l$ along with token routers and $\nn$ in Fig. \ref{fig: keep_ratio}. For a given token input $\multimodalin$, we first round $\tokentotalnum\keepratio_l$ to the two nearest integers $\topk_l^\text{up}$ and  $\topk_l^\text{down}$. Then, we do two forward passes with Top-$\topk_l^\text{up}$ and Top-$\topk_l^\text{down}$ tokens, respectively. However, Top-$\topk$ operation is not differentiable, making it challenging to directly optimize $\keepratio_l$. Hence, we combine the outputs of the two forward passes as follows
\begin{align}
	\bar{\multimodalin} \leftarrow \bar{\multimodalin}^{\text{up}} \times (\tokentotalnum\keepratio_l - \topk_l^\text{down}) + \bar{\multimodalin}^{\text{down}} \times (\topk_l^\text{up} - \tokentotalnum\keepratio_l), \label{keepratio_grad}
\end{align} 
where $\bar{\multimodalin}^{\text{up}}$ and $\bar{\multimodalin}^{\text{down}}$ are the outputs of forward passes with Top-$\topk_l^\text{up}$ and Top-$\topk_l^\text{down}$ as in \eqref{router_grad}, respectively. We then can create gradient paths to $\keepratio_l$ by \eqref{keepratio_grad}. Meanwhile, prior works \cite{10171192, yinunderstanding, jang2017categorical} often used either straight-through estimators (STE) \cite{yinunderstanding, 10171192} or Gumbel Softmax relaxations \cite{jang2017categorical} to address discrete operators (e.g., Top-$\topk$, quantization or pruning).   However, STE uses surrogate gradients that are not the true gradients of forward propagation and can suffer from unstable training \cite{yinunderstanding}. Moreover, Gumbel Softmax relaxations often introduce a temperature hyperparameter, which can lead to high variance and unstable gradients when the temperature in not appropriately controlled \cite{jang2017categorical}. In contrast, our approach generates gradients of the relaxed objective through piecewise relaxation by interpolating two forward outputs with respect to keep ratios $\keepratio_l, \forall l$, rather than yielding approximated gradients through Top-$\topk$.

To control the total amount of tokens to process, we define a target keep ratio $\flopbound' \in [0, 1]$. For instance, if $\flopbound' = 0.3$, we aim to process only 30\% of  tokens to satisfy a desired computational budget\footnote{To enforce \eqref{constraint_1},  we can set $ \flopbound' = \sqrt{(\gamma/\text{FLOPs}_\text{max})}$, where $\text{FLOPs}_\text{max}$ is the maximum \ac{FLOPs}. For keep ratio $\keepratio_l$, the \ac{FLOPs} of the attention can be approximated by $\text{FLOPs}(\topk_l) \approx \topk_l^2 \tokyodimension$. Then, we can derive the normalized \ac{FLOPs} ratio as $\frac{\text{FLOPs}(\topk_l)}{\text{FLOPs}(\tokentotalnum)} \approx (\frac{\topk_l}{\tokentotalnum})^2 = \keepratio_l^2$. Thus, $\flopbound' = \sqrt{\gamma/\text{FLOPs}_{\text{max}}}$ can provide a soft penalty term for keep ratios to enforce $\flopbound$.  Essentially, $\flopbound'$ is the target fraction of tokens to satisfy the normalized \ac{FLOPs} budget.}\footnote{Dynamically adjusting $\flopbound'$ based on available hardware resources and energy constraints can be an important subject of future research for dynamic IoT environments.}. To this end, we add an \ac{MSE} loss function $(\keepratio_\text{avg} - \flopbound')^2$, where $\keepratio_\text{avg} = \frac{1}{\tokyolayernum} \sum_l^\tokyolayernum \keepratio_l$. Here, $\keepratio_\text{avg}$ is not an exact \ac{FLOPs} estimator. However, $\keepratio_\text{avg}$ still provides a reasonable estimate because the \ac{FLOPs} of attention scales quadratically with respect to the number of tokens as $\mathcal{O}(\topk_l^2 \tokyodimension)$, and $\topk_l = \lceil \keepratio_l\tokentotalnum \rceil$. As such, decreasing $\keepratio_l$ monotonically decreases $\topk_l$, reducing attention \ac{FLOPs}.  Hence, we use $\keepratio_\text{avg}$ as a global and differentiable variable to reduce $\topk_l$ through optimized $\keepratio_l$ across encoder blocks. Hence, the total loss function is as follows
\begin{align}
	\tokyoloss_\text{total}(\nn, \boldsymbol{\topk})  =\tokyoloss(\nn, \boldsymbol{\topk}) + \lambda (\keepratio_\text{avg} - \flopbound')^2 \label{final_loss}
\end{align}
where $\lambda$ is a hyperparameter to enforce \eqref{constraint_1}. Hence, \eqref{final_loss} is a relaxed problem of \eqref{problem}. Every encoder block is architecturally identical including embedding size, number of heads, and linear layers. Hence, each encoder block has the same computational costs at the beginning of the training. However, each encoder block eventually receives different gradients because \eqref{final_loss} has both the performance loss function and soft penalty term. Therefore, each $\keepratio_l, \forall l$ will be automatically optimized based on performance contribution and efficiency regularization.

We now discuss the convergence aspect of $\tokyoloss_\text{total}(\nn, \boldsymbol{\topk})$ by following the framework in \cite{lei2019stochastic}. For notational simplicity, we use $\tokyoloss(\nn)$ for $\tokyoloss_\text{total}(\nn, \boldsymbol{\topk})$ and define $\xi(\nn) = \E_{\multimodalin, \out}[\tokyoloss(\nn)]$, where $\multimodalin$ and $\out$ are input and ground truth, respectively. Note that $\tokyoloss(\nn)$ is generally nonconvex. Hence, we assume that the gradient of $\tokyoloss(\nn)$ is $\alpha$-Holder continuous such that $||\nabla\tokyoloss(\nn) - \nabla\tokyoloss(\nn')||_2 \leq L ||\nn - \nn'||^\alpha_2, \forall \nn, \nn'$, where $\alpha \in (0,1]$ and $L >0$. Essentially,  $\alpha$-Holder continuous is a generalized version of $L$-smoothness condition, which is widely used for convergence analysis of gradient based methods \cite{10171192,lei2019stochastic}. Assume that we update our model $\nn$ using stochastic gradient descent to minimize $\tokyoloss(\nn)$. Then, for iteration $t$ with learning rate $\eta_t$, the update rule can be given as $\nn_{t+1} = \nn_t - \eta_t \nabla(\tokyoloss(\nn_t))$. Then, according to \cite{lei2019stochastic}, under the assumption of the $\alpha$-Holder continuous and learning rates satisfying $\sum_{t=1}^\infty \eta_t^{1+\alpha} < \infty$, there is a constant $C$ independent of $t$ such that 
	\begin{align}
		\min_{t=1,...,T} \E \big[ || \nabla \xi(\nn_t)||^2_2   \big] \leq C \big( \sum_{t=1}^\infty \eta_t \big)^{-1}.
	\end{align}
	Hence, our model update rule satisfies an asymptotic bound on $\min_{t} \E \big[ || \nabla \xi(\nn_t)||^2_2$.

\subsection{Complexity Analysis} \label{sub: complexity}
We now provide a systematic analysis of computational and memory costs of the proposed framework. We first analyze the complexity of tokenizers, routers, and attention modules of one inference.
For image and LiDAR samples, we use a ResNet-based CNN blocks. From \cite{kim2024spafl}, the computational complexity of convolutional layer $j$ can be given by $\mathcal{O}(H_j W_j k_j^2 F_j C_{j})$, where $H_j \times W_j$ is the resolution of the input in layer $j$, $k_j$ is the kernel size and $C_j$ is the number of channels, and $F_j$ is the number of filters. Then, the corresponding memory cost can be given by $\mathcal{O}(H_j^2 \times F_j + W_j^2 \times C_j)$ \cite{lin2021memory}. For radar, GPS, and \ac{RSSI}, we use a few linear layers. The computational complexity of linear layer $j$ can be given by $\mathcal{O}(I_j \times O_j) $, where $I_j$ and $O_j$ are input and output dimensions, respectively \cite{kim2024spafl}. Then, the corresponding memory consumption can be derived as $\mathcal{O}(I_j + O_j)$. Routers consist of linear layers and Top-$K$ selections. Hence, for $\tokentotalnum$ and $\topk_j$ selected tokens in router $j$, the computational complexity can be given by $\mathcal{O}(\tokentotalnum \tokyodimension^2 + \tokentotalnum \log \topk_j)$. The corresponding memory cost can be given by $\mathcal{O}(\topk_j \tokyodimension + \tokyodimension)$ by indexing and storing selected tokens. For encoder block $j$, the computing complexity of attention is $\mathcal{O} (\topk_j^2\tokyodimension)$. The corresponding memory cost of the attention module will be $\mathcal{O} (\topk_j \tokyodimension)$.

The keep-ratio optimization essentially does two-forward propagations within encoder blocks during the training. Hence, it doubles the computational and memory costs of one forward propagation of encoder blocks. However, note that our main focus is to improve the efficiency of multli-modal transformer \emph{inference} for wireless communication tasks. Moreover, the keep-ratio optimization can decrease the latency of one training epoch because we still do Top-$K$ operations in encoder blocks. Consider a baseline which process every token $\tokentotalnum$ in each encoder block. Then, the computational complexity can be given by $\mathcal{O}(\tokentotalnum^2)$, where $\tokentotalnum$ is the total number of tokens. Essentially, the computational complexity of the keep-ratio optimization can be approximated as $\mathcal{O}(2\topk^2)$. Hence, we can still improve the computational complexity of the training when $\topk < \frac{\tokentotalnum}{\sqrt{2}}$. We visualize this observation in Fig. \ref{fig: training_epoch}.

In summary, the proposed framework uses tokenizers and token-level routers to learn the importance of tokens in the same embedding dimension. Then, trainable keep ratios are defined to optimize $\boldsymbol{K}$ with a target ratio $\flopbound'$ to control the inference latency. Here, $\flopbound'$ can be set as a proxy variable for the target latency, such as beam coherence time, to support the current wireless communication tasks. Next, we validate the proposed framework using 1) public multi-modal wireless communication datasets and 2) our measurement-based real-world datasets.

\section{Simulation Results and Analysis} \label{sec: tokyo_experiments}
We evaluate our framework on two public multi-modal mmWave beamforming task datasets \cite{alkhateeb2023deepsense, park2025resource} and one real-world dataset that we collected in an actual testbed for the purpose of executing a multi-modal proactive mmWave handover task.

\subsection{mmWave Beamforming Tasks} \label{subsec: beamforming}
We first consider the use of multi-modal data for mmWave beamforming tasks \eqref{prob: beamforming} in vehicular networks. Here, multi-modal inputs are beneficial for mmWave beamforming tasks because high frequency channels are susceptible to mobility and blockages, necessitating situational-awareness. We use the following two public datasets: (1) \textbf{Scenarios 31-34 from DeepSense 6G} \cite{alkhateeb2023deepsense} which predict 64 beamforming indices based on the five sequence of image, LiDAR, radar, and GPS inputs from real-world setup; (2) \textbf{CARLA-MATLAB-based simulation dataset} \cite{park2025resource} that is used to predict 152 beamforming indices based on the five sequence of image, LiDAR, radar, and GPS inputs from simulations. In \cite{park2025resource}, the optimal beam labels maximize the average received signal strength of multiple vehicles that are captured in multi-modal sensors for given frames. For the image and LiDAR tokenizers, we use the first four blocks of ResNet-18 and a 1-D convolutional layer with the output embedding dimension $\tokyodimension = 64$. The resolution of each image of the both datasets is $960\times540$. To keep the detailed spatial information of LiDAR data, we set the resolution of the corresponding bird's eye view representation as $512\times512$. We generate $1024$ tokens for each image and LiDAR sample following  \cite{team2024chameleon,bergner2025token}. For the radar and GPS tokenizers, we use two linear layers with the same output embedding dimension $\tokyodimension = 64$. We generate one token for each radar cube and set the number of radar cubes $N_\text{rad}=300$. We assigned one token for GPS. Then, the input sequence length of five frames of image, LiDAR, radar, and GPS will be $\tokentotalnum= 11745$.  We use 8 layers of a transformer block, which has multi-head-self-attention with 8 heads, two linear layers, layer norm, and a token router. We use a single linear layer for the head to make predictions. We use position embedding nn.Parameter(torch.zeros(1, 32 $\times$ 32, d)) for each image and LiDAR sample. We also use time embedding  nn.Embedding(5, d) for each time frame of the input sequence, and modality embedding nn.Embedding(4, d) for each modality. We add one [CLS] token to the concatenated token input $\multimodalin$ for the prediction. 

We split the datasets into training and validation as 90-10\% split following \cite{tian2023multimodal, shang2025multi, 10735366}.  We set the learning rate as $0.0001$, the weight decay as $0.01$, and the batch size as 16 for 30 epochs. We use $\lambda = 10$ for the computational constraint \eqref{constraint_1}. We use an ADAM optimizer and average every result over five random seeds. We use NVIDIA A30 for our simulations. We consider three baselines to compare the performance and efficiency. \textbf{Baseline [MHA]} uses the same architecture but it process every token with multi-head attention (MHA) similar to \cite{cui2024sensing}, where the authors forwarded every modality token to the same transformer blocks. \textbf{Baseline [ToMe]}  \cite{bolya2023token} merges fixed number of similar tokens in each transformer encoder block. Baseline [ToMe] was proposed for image tasks, but we modify the algorithm to our multi-modal wireless communications scenarios. \textbf{Baseline [MoD]} \cite{mod} uses a token router to process only Top-$K$ tokens based on measured importance. The solution in \cite{mod} derives $K$ heuristically by setting a token router in every other transformer encoder block. Hence, we set layers with odd indices process every token and layers with even indices process Top-$K$ tokens based on their implementation \cite{mod}. Since \cite{mod} was designed for language tasks, we modify the algorithm for wireless multi-modal scenarios. We set Baseline [ToMe] and Baseline [MoD] to process around 30\% and 55\% of tokens, respectively, by following the official and recommended configuration in \cite{bolya2023token} and \cite{mod}.

\begin{table*}[t!] 
	\centering
	\begin{tabular}{|c|c|c|c|c|c|c|}
		\hline
		\textbf{Methods} &
		\textbf{Top-1 {[}\%{]}} &
		\textbf{Top-3 {[}\%{]}} &
		\textbf{Top-5 {[}\%{]}} &
		\textbf{Latency {[}ms{]}} &
		\textbf{Memory {[}GB{]}} &
		\textbf{FLOPs {[}G{]}} \\ \hline
		Baseline [MHA] & $47.5\pm0.9$  & $80.9\pm1.0$  & $91.73\pm0.5 $& 158.14 & 0.797 &  292.19 \\ \hline
		Baseline [ToMe] & $46.2\pm1.5$  & $79.5\pm	0.9$  & $90.5\pm0.9 $& 127.77 & 0.615 &  160.17 \\ \hline
		Baseline [MoD] & $46.78\pm0.7$  & $80.82\pm	0.7$  & $91.6\pm0.3 $& 118.82 & 0.622 &  154.10 \\ \hline
		Proposed ($\flopbound' =30\%$) & $47.38\pm1.3$ & $81.32\pm1.3$ & $91.16\pm0.8$ & \textbf{21.70}  & \textbf{0.511} & \textbf{28.60} \\ \hline
		Proposed ($\flopbound' = 40\%$) & $47\pm1.1$    & $81.74\pm0.8$ & $91.12\pm0.3$ & 33.09  & 0.559 & 49.31 \\ \hline
		Proposed ($\flopbound' = 50\%$) & $\boldsymbol{47.84\pm0.9}$ & $81.8\pm0.8$  & $92.08\pm0.6$ & 46.81  & 0.592 & 75.65  \\ 	 \hline
		Proposed ($\flopbound' = 70\%$) & $47.8\pm0.6$  & $\boldsymbol{82.46\pm0.9}$ & $\boldsymbol{92.1\pm0.5}$  & 81.79  & 0.676 & 145.34 \\ \hline
	\end{tabular}
	\caption{Performance, inference latency, GPU memory consumption, and FLOPs on DeepSense 6G scenarios 31-34 \cite{alkhateeb2023deepsense}.} 
	\label{tab: d6g}
\end{table*}

\begin{table*}[t!] 
	\centering
	\begin{tabular}{|c|c|c|c|c|c|c|}
		\hline
		\textbf{Methods} &
		\textbf{Top-1 {[}\%{]}} &
		\textbf{Top-3 {[}\%{]}} &
		\textbf{Top-5 {[}\%{]}} &
		\textbf{Latency {[}ms{]}} &
		\textbf{Memory {[}GB{]}} &
		\textbf{FLOPs {[}G{]}} \\ \hline
		Baseline [MHA] & $\boldsymbol{27.76\pm1.5}$  & $\boldsymbol{49.87\pm1.3}$  & $\boldsymbol{62.45\pm1.6} $& 158.14 & 0.797 &  292.19 \\ \hline
		Baseline [ToMe] & $25.62\pm1.4$  & $47.30\pm1.2$  & $59.85\pm0.9 $& 127.77 & 0.615 &  160.17 \\ \hline
		Baseline [MoD] & $26.00\pm1.8$  & $45.94\pm2.8$  & $60.2\pm1.8 $& 118.81 & 0.622 &  154.10 \\ \hline
		Proposed ($\flopbound' =30\%$) & $26.27\pm0.8$ & $48.2\pm1.5$ & $60.86\pm1.3$ & \textbf{27.18}  & \textbf{0.516} & \textbf{28.60} \\ \hline
		Proposed ($\flopbound' =40\%$) & $25.47\pm0.8$    & $47.74\pm0.9$ & $60.1\pm1.0$ & 33.24  & 0.565 & 49.31 \\ \hline
		Proposed ($\flopbound' =50\%$) & $26.54\pm0.7$ & $48.5\pm1.1$  & $61.5\pm1.4$ & 46.90  & 0.594 & 75.65  \\ 	 \hline
		Proposed ($\flopbound' =70\%$) & $27.2\pm0.6$  & $49.2\pm1.5$ & $62.2\pm1.4$  & 81.95  & 0.684 & 145.34 \\ \hline
	\end{tabular}
	\caption{Performance, inference latency, GPU memory consumption, and FLOPs on \cite{park2025resource}.} 
	\label{tab: yumin}
\end{table*}

Table \ref{tab: d6g} shows the top-1, 3, and 5 accuracy on the DeepSense 6G datasets \cite{alkhateeb2023deepsense} with the inference latency, GPU memory, and FLOPs for processing batch size of one. We can see that our proposed approach can significantly reduce the inference latency compared to Baseline [MHA]. For $\flopbound' = 30\%$, which targets processing 30\% of tokens, we can improve the latency by 86.2\% with only $0.12\%$ of top-1 accuracy loss. We can also observe that we can improve the GPU memory usage and FLOPs by 35\% and 80\%, respectively. We can also see that the complexity of our framework does not increase for processing multiple vehicles compared to \cite{alkhateeb2023deepsense}. This is because the complexity of our framework does not scale with the number of vehicles as long as we only consider vehicles captured in multi-modal sensors in given frames. \footnote{However, if we install more sensors (e.g., more cameras to capture another angle) to extend the covered area, then the number of input tokens increases accordingly. Then, the complexity of the Top-$\topk$ selection increases linearly with respect to the number of input tokens as discussed in Section. \ref{sub: complexity}.} Meanwhile, for $\flopbound' = 50\%$ and $\flopbound' = 70\%$, we can see that the proposed framework achieves better accuracy than Baseline [MHA]. We conjecture that the token routers provide regularization effects because they make the model focus on only important tokens while skipping redundant tokens. Baseline [ToMe] decreases latency less than our method at $\flopbound'=50\%$. This is because Baseline [ToMe] computes the pairwise similarities between tokens in the two subsets, resulting in quadratic complexity with respect to the number of tokens in each subset \cite{bolya2023token}. Moreover, it merges tokens after attention modules.  Hence, Baseline [ToMe] has additional overhead for merging tokens as shown in \cite{gotz2025efficient}. For $\flopbound' = 50\%$, we can improve the top-1 accuracy, latency, GPU memory, and FLOPs by 1.64\%, 65\%, 10\%, 58\% compared to Baseline [ToMe]. We can observe that Baseline [MoD] has lower accuracy and higher latency, and FLOPs than our $\flopbound' = 70\%$. Baseline [MoD] has token routers in every other transformer blocks based on heuristic design \cite{mod}, thereby leading to sub-optimal performance. 

In Table \ref{tab: yumin}, we present the top-1, 3, and 5 accuracy on the datasets from \cite{park2025resource} with the inference latency, GPU memory, and FLOPs for processing a batch size of one. We can see that, for $\flopbound' = 50\%$, we can improve the latency, memory, and FLOPs by 70.3\%, 25.4\%, and 74.1\%, respectively, with only 1.22\% top-1 accuracy loss compared to Baseline [MHA]. For $\flopbound' = 70\%$, we can still improve the latency, memory, and FLOPs by 48.2\%, 14.2\%, and 50.2\%, respectively, with only 0.56\% top-1 accuracy loss. Moreover, we can improve the top-1 accuracy by 0.92 compared to Baseline [ToMe]. Note that each time frame in \cite{park2025resource} is 100 ms. As such, Baselines [MHA] and  [ToMe] cannot predict a beamforming index within the time frame, thereby making obsolete predictions. For Baseline [MoD], we can improve the top-1 accuracy, latency, and FLOPs by 0.4, 62\%, and 51\% compared to our $\flopbound' = 50\%$. 

\begin{figure}[t]
	\centering 				%
	\begin{subfigure}[t]{0.49\linewidth}
		\centering
		\includegraphics[width=\textwidth]{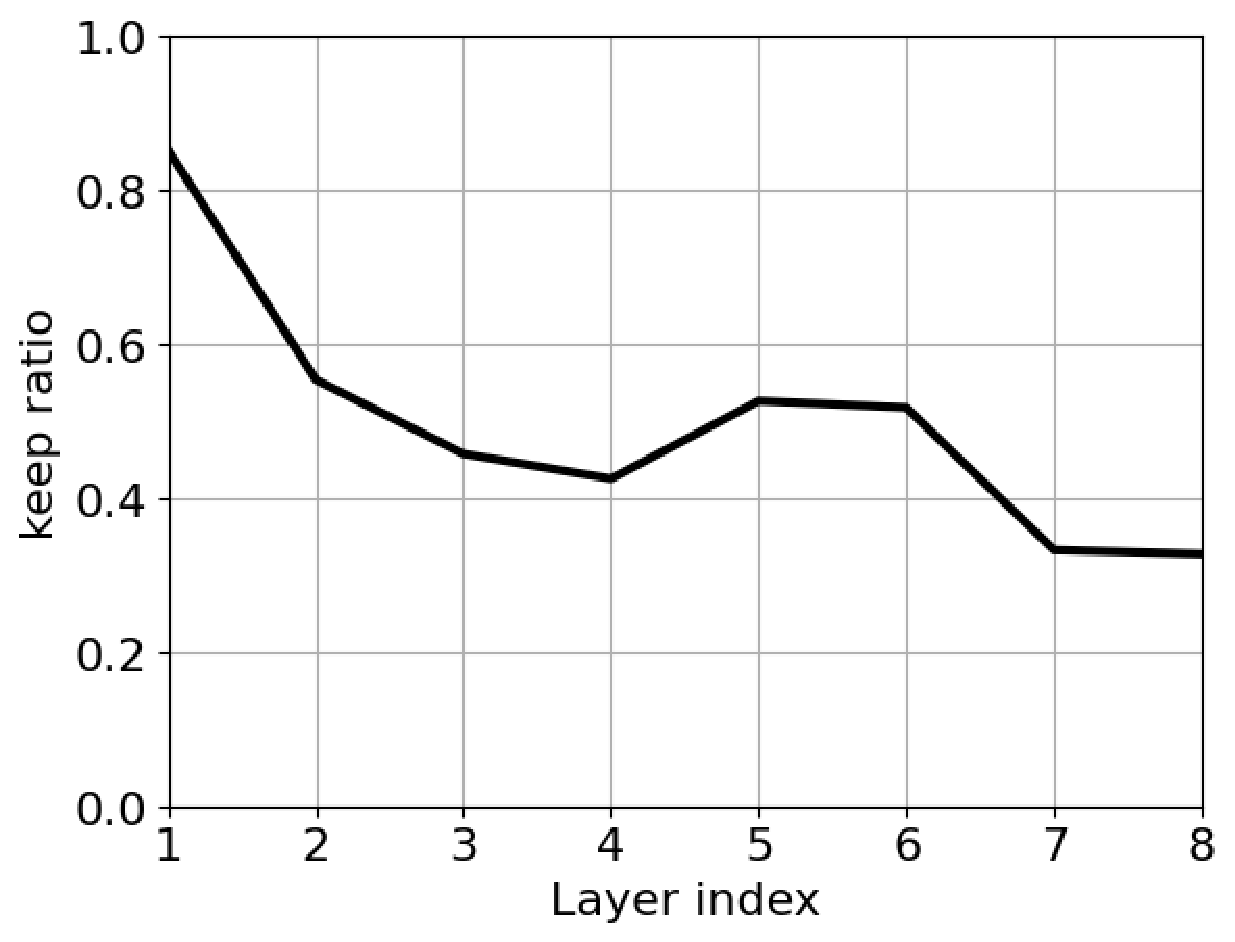}
		\caption{Optimized keep ratios for the DeepSense 6G datasets.}
	\end{subfigure}
	\begin{subfigure}[t]{0.49\linewidth}
		\centering
		\includegraphics[width=\textwidth]{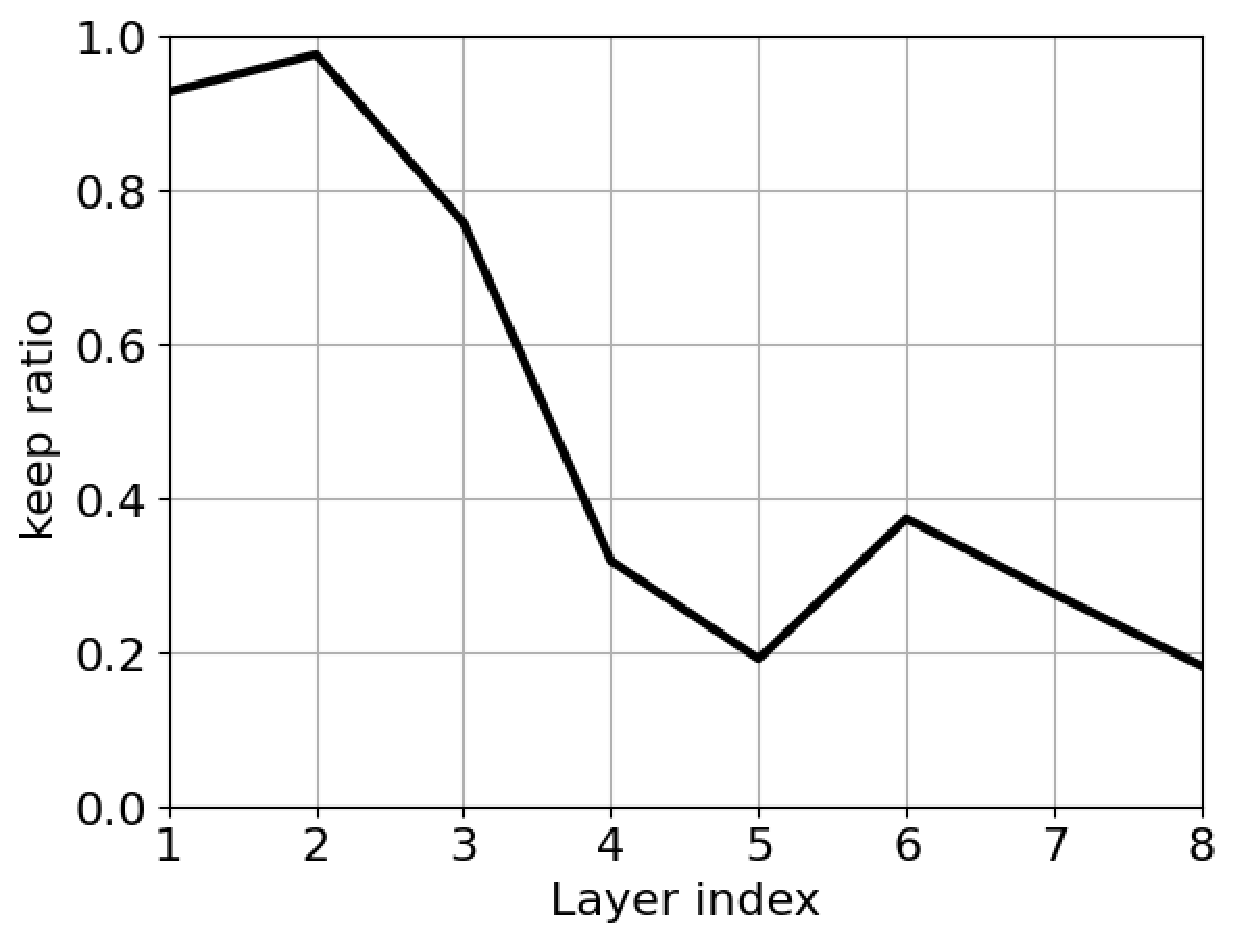}
		\caption{Optimized keep ratios for the datasets in \cite{park2025resource}.}
	\end{subfigure}
	
	\caption{Optimized keep ratios across transformer encoder blocks}
	\label{fig: keep_ratios}
\end{figure}

Figure \ref{fig: keep_ratios} shows the learned keep ratios across transformer encoder blocks on the mmWave beamforming datasets for $\flopbound' = 50\%$. We can see that optimized keep ratios are not uniform across layers. In general, the optimized keep ratios decrease as the depth of models increases. Earlier layers often extract more important features than back layers as experimentally proven in \cite{kim2024spafl}. As such, the front layers need to process more tokens than the back layers. 

In Tables \ref{tabl: ablation_d6g} and \ref{tabl: ablation_yumin}, we present the result of an ablation study on the proposed framework with datasets from DeepSense 6G and \cite{park2025resource} for $\flopbound' = 30\%$. We evaluate the impact of each main component on the accuracy.  Ablation 1 freezes the tokenizers which results in large modality gap between tokens from different modalities. Ablation 2 freezes token routers and processes tokens randomly. Ablation 3 uniformly processes $30\%$ of tokens across all transformer blocks and does not train keep ratios $\keepratio$. We can see the importance of tokenizers from Ablation 1. The performance drops significantly because tokens are not embedded to be in the same space, thereby resulting in large modality gaps.  Ablation 2 shows the accuracy gain from token routers. Important tokens can be skipped because tokens routers are frozen in Ablation 2. We can see the importance of optimizing keep ratios from Ablation 3. Each layer needs different optimal keep ratio $\keepratio$ as shown in Fig. \ref{fig: keep_ratios}. Hence, an optimal $\keepratio$ should be found during training rather than setting heuristic rules. 

\begin{table}[]
	\centering
	\begin{tabular}{|c|c|c|c|}
		\hline
		\textbf{Methods} & \textbf{Top-1  {[}\%{]}} & \textbf{Top-3 {[}\%{]}} & \textbf{Top-5 {[}\%{]}} \\ \hline
		Proposed ($\flopbound' = 30\%$)  & $47.38\pm1.3$ & $81.32\pm1.3$ & $91.16\pm0.8$ \\ \hline
		Ablation 1 & $15.9\pm1.2$  & $31.6\pm1.9$  & $44\pm3.2$    \\ \hline
		Ablation 2 & $44.6\pm0.7$  & $77.7\pm1.0$  & $89.3\pm0.7$  \\ \hline
		Ablation 3 & $40.1\pm2$    & $71.1\pm3.5$  & $84.4\pm2.5$  \\ \hline
	\end{tabular}
	
	\caption{Ablation study on DeepSense 6G with $\flopbound' = 30\%$}
	\label{tabl: ablation_d6g}
\end{table}

\begin{table}[]
	\centering
	\begin{tabular}{|c|c|c|c|}
		\hline
		\textbf{Methods} & \textbf{Top-1  {[}\%{]}} & \textbf{Top-3 {[}\%{]}} & \textbf{Top-5 {[}\%{]}} \\ \hline
		Proposed ($\flopbound' = 30\%$)  & $26.27\pm0.8$ & $48.2\pm1.5$ & $60.86\pm1.3$ \\ \hline
		Ablation 1 & $20.2\pm1.3$  & $40.8\pm0.7$  & $51.8\pm2.3$    \\ \hline
		Ablation 2 & $23.7\pm2.0$  & $45.4\pm1.0$  & $57.2\pm2.0$  \\ \hline
		Ablation 3 & $22.7\pm1.9$    & $44\pm1.4$  & $54.2\pm2.2$  \\ \hline
	\end{tabular}
	
	\caption{Ablation study on \cite{park2025resource} with $\flopbound' = 30\%$}
	\label{tabl: ablation_yumin}
\end{table}

\subsection{Proactive mmWave Handover with Real-World Dataset}
\begin{figure*}[t]
	\centering
	\begin{subfigure}[t]{0.49\linewidth} 				%
	\includegraphics[width=\textwidth]{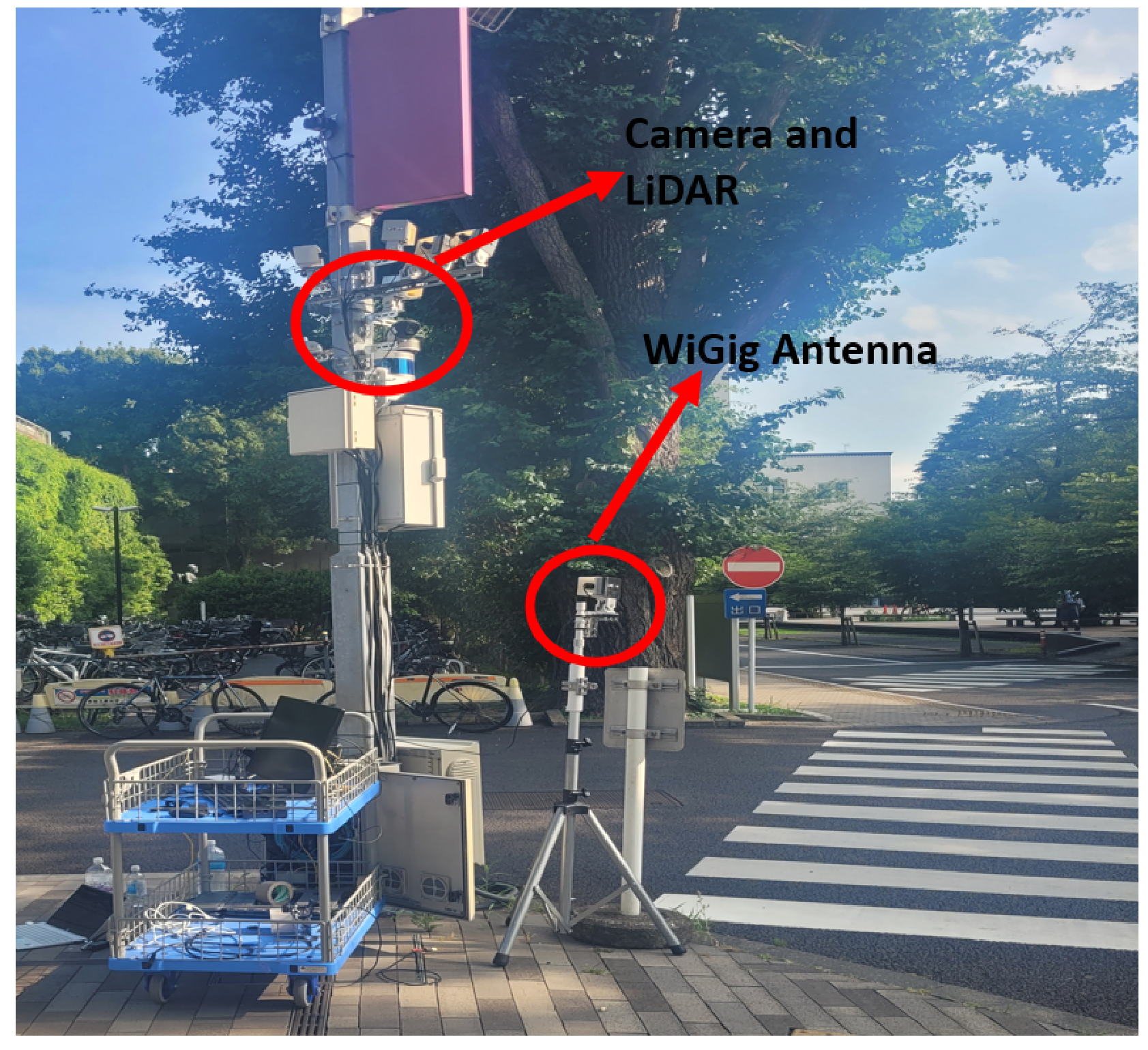}
	\caption{Setup of the testbed for multi-modal proactive mmWave handover.}
	\label{fig: testbed}
	\end{subfigure}
    \hfill
   	\begin{subfigure}[t]{0.49\linewidth}
   	\includegraphics[width=\textwidth]{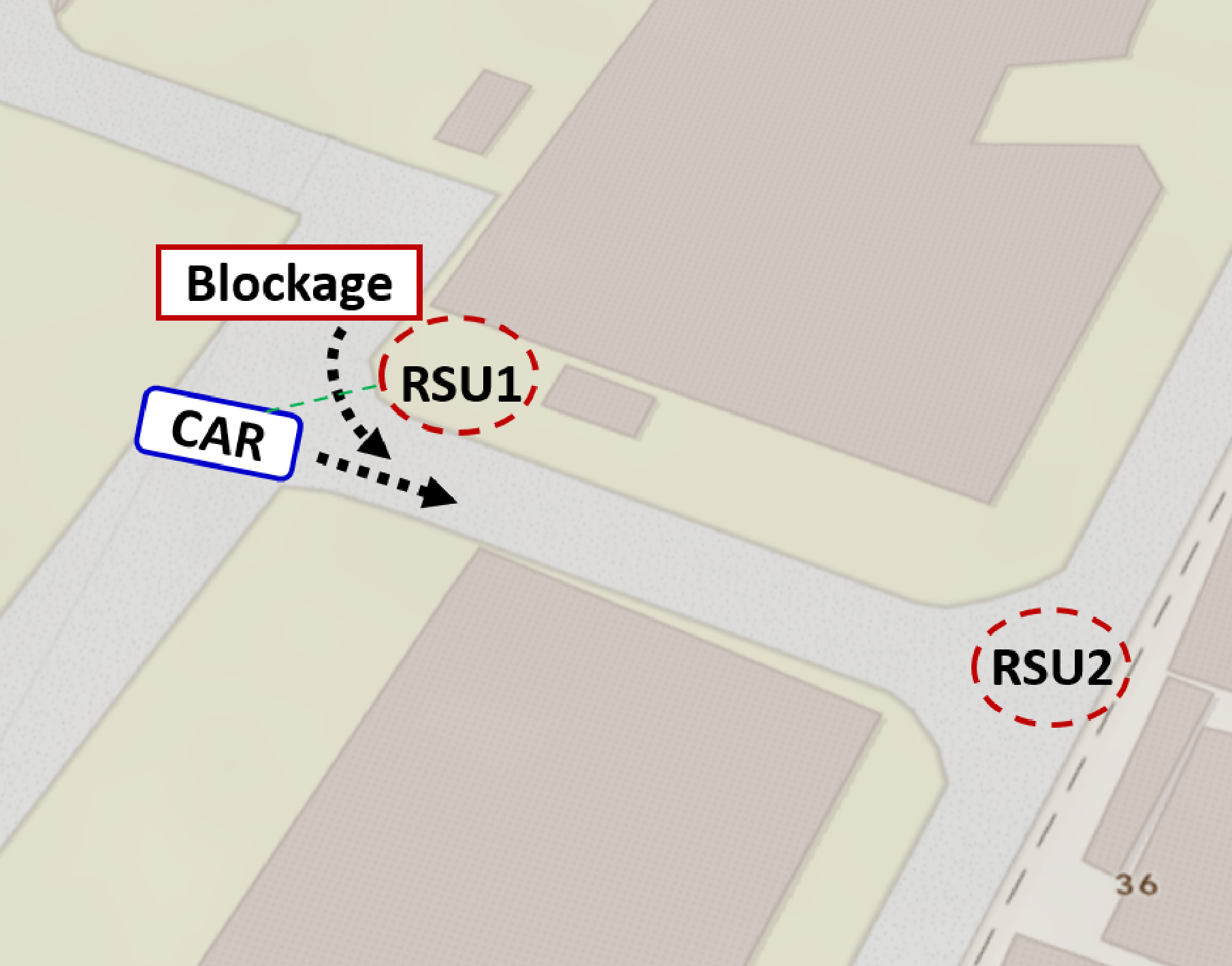}
   	\caption{Illustration of the testbed, \ac{RSU}s, and vehicles.}
   	\label{fig: testbed_map}
   \end{subfigure}
   \caption{Overview of the multi-modal mmWave handover testbed and setup.}
   \label{fig: setup}
\end{figure*}

To further validate our proposed framework on real-world datasets, we collect data from a testbed located at the Institute of Science Tokyo, Ookayama campus, Japan, as shown in Fig. \ref{fig: testbed}. We use one vehicle and two \ac{RSU}s for the data collection as shown in Fig. \ref{fig: testbed_map}. Specifically, we mount a WiGig 60 GHz antenna on the vehicle, and install the same antenna on each \ac{RSU}. We equip each \ac{RSU} with a fixed 80-layer 3D LiDAR (RS-LiDAR 80) \cite{lidar} and a camera. The resolution of a captured image is  $640 \times 480$. We use the same setting for LiDAR and embeddings as done previously with the beamforming datasets. We generate $1024$ tokens for each image and LiDAR frame and one token for each \ac{RSSI}. Then, the input sequence of five frames of image, LiDAR, and \ac{RSSI} will be $\tokentotalnum = 11446$ with the CLS token.  We drove the vehicle at the maximum speed of 10 km per hour from \ac{RSU} 1 to \ac{RSU} 2 by collecting \ac{RSSI} between the car and the \ac{RSU}s every 300 ms. Simultaneously, \ac{RSU} 1 collects point clouds and image data at every 300 ms. We use a large vehicle to block the mmWave link between the target vehicle and \ac{RSU} 1. We synchronize RSSI, LiDAR, and image within the average of $50$ ms. To the best of our knowledge, there is no public dataset for proactive mmWave handover based on LiDAR, image, and \ac{RSSI}.

\begin{table*}[]
	\centering
	\begin{tabular}{|c|c|c|c|c|}
		\hline
		\textbf{Methods} & \textbf{Accuracy [\%]} & \textbf{Latency. [ms]} & \textbf{Memory [GB]} & \textbf{FLOPs [G]} \\ \hline
		Baseline [MHA]   & $94.30\pm4.7$          & 335.93         & 1.292          &151.12  \\
		Baseline [ToMe]   & $92.39\pm4.8$          & 285.54         & 1.205         &85.87  \\
		Baseline [MoD]   & $93.47\pm4.6$          & 197.72        & 1.292          &79.56 \\
		Proposed ($\flopbound' =30\%$) & $92.73\pm4.3$         & \textbf{56.84} & \textbf{1.014} & \textbf{19.23}  \\
		Proposed ($\flopbound' =40\%$) & $93.69\pm5.1$         & 77.07 & 1.074 & 29.67  \\
		Proposed ($\flopbound' =50\%$) & $94.00\pm4.9$         & 101.83 & 1.184 & 42.85  \\
		Proposed ($\flopbound' =70\%$) & $\boldsymbol{94.49\pm4.8}$ & 170.22         & 1.303          &77.69  \\ \hline
	\end{tabular}
	\caption{Performance, inference latency, GPU memory consumption, and FLOPs on  our collected data.}
	\label{tab: blockage}
\end{table*}

We collected around 800 data samples over 10 scenarios.\footnote{We note that our measurement scale is relatively small and collected over 10 scenarios in a single site. Expanding the testbed (more locations, traffic densities, vehicle types/speeds, and blockage sources) is an important direction for future work  } We label a data sample as `blocked' if the measured \ac{RSSI} between the vehicle and \ac{RSU} 1 is below -47 dBm. Hence, we set $\tokyorss_{\text{th}} = -47$ as a condition to trigger handover \eqref{prob: handover}. We group the past five data samples for as an input and predict the blockage status of the next time frame. Hence, we assume a scenario which decides handover execution for the next 300 ms, given the past five data frames. We use the same model architecture as in Sec. \ref{subsec: beamforming} with four transformer blocks. We use the same hyperparameters as in Sec. \ref{subsec: beamforming}, except that we set epochs as 20. We report the averaged accuracy with 9-fold cross-validation over 10 scenarios. We average every result over five random seeds. In this scenario, we emulate the handover execution. Specifically, if \ac{RSU} 1 predicts the blockage successfully within a time frame we report the \ac{RSSI} between the vehicle and \ac{RSU} 2. \ac{RSU}s often have limited computing capability and memory. To capture this setting, we benchmark inference latency on an NVIDIA T4 GPU, which is widely adopted for other works that considered edge scenarios \cite{carlak2024state, yang2024navigator, zhang2021real}.

In Table \ref{tab: blockage}, we show the accuracy on our collected datasets with the inference latency, GPU memory, and FLOPs for processing batch size of one. We can see that, for $\flopbound' = 40\%$, we can improve the latency, GPU memory, and FLOPs by 83.2\%, 16.9\%, and 80\%, respectively, with only 0.61\% accuracy loss. We also observe that, for $\flopbound' = 70\%$, the latency, GPU memory, and FLOPs  can be improved by 49.3\%, 8.43\%, and 49.2\%, respectively, compared to the baseline. Moreover, our proposed method with $\flopbound'=50\%$ outperforms Baseline [ToMe]  in accuracy, latency, and FLOPs by 1.61\%, 64\%, and 63\%, respectively. Baseline [ToMe] has limited efficiency gain due to the additional overhead for merging tokens, and it reduces the number of tokens after attention. Although Baseline [MoD] outperforms Baseline [ToMe], its accuracy is lower than our $\flopbound' = 70\%$ because it sets keep ratios heuristically. 

\begin{figure*}[t!]
	\begin{subfigure}[t]{0.49\textwidth}
		\begin{center}   				%
			\includegraphics[width=\textwidth]{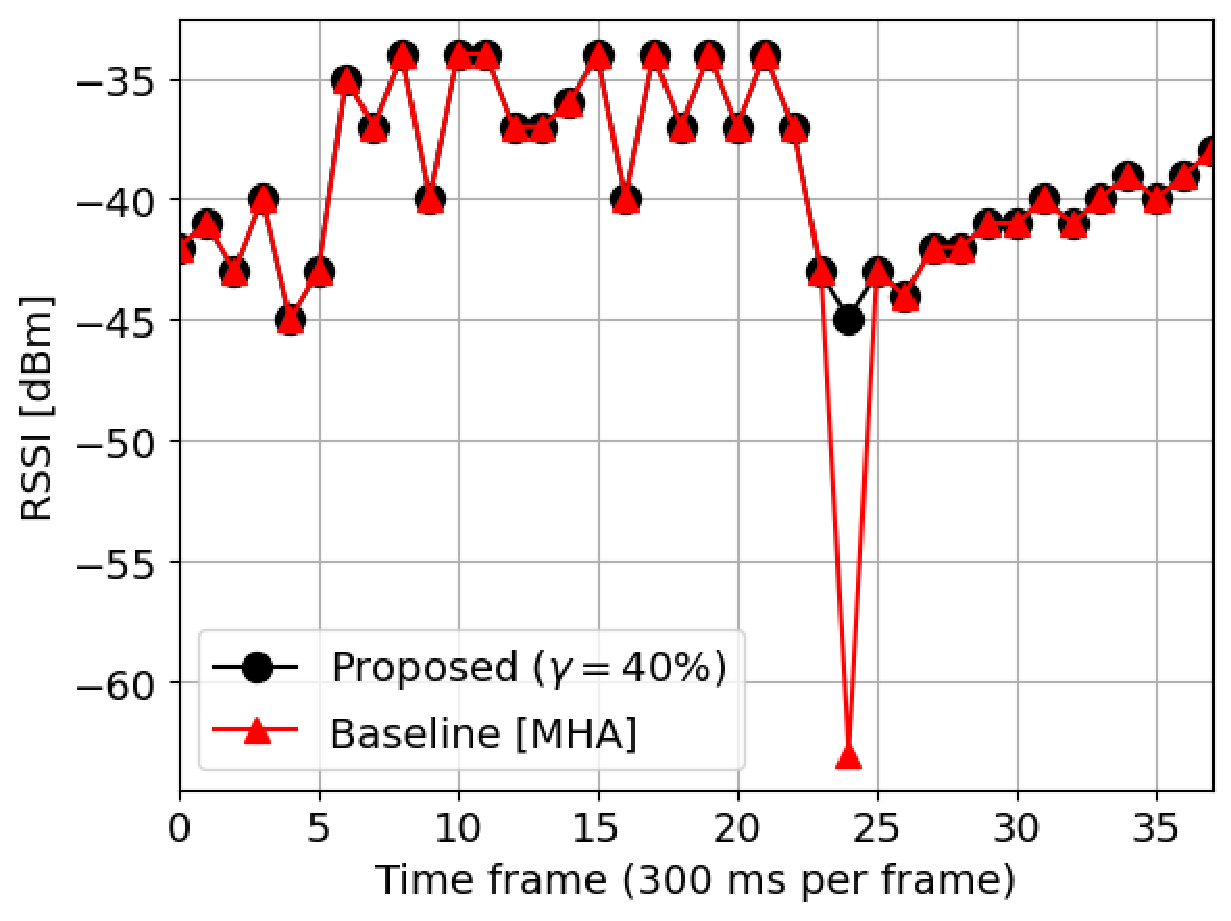}
		\end{center}
		\caption{Change of \ac{RSSI} with the blockage and proactive handover in Scenario 5.}
		\label{fig: scenario5}
	\end{subfigure}\hfill
	\begin{subfigure}[t]{0.49\textwidth}
		\begin{center}   
			\includegraphics[width=\textwidth]{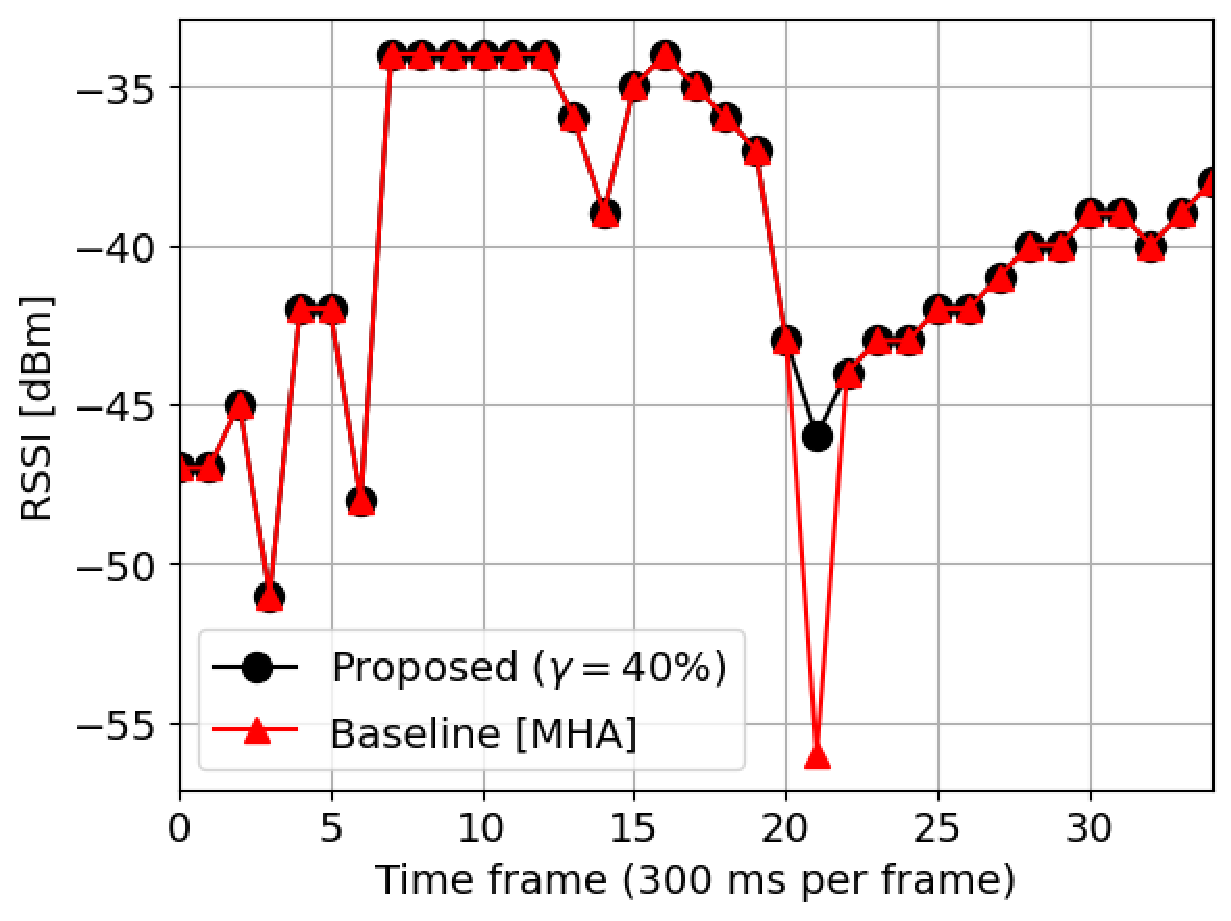}
		\end{center}
		\caption{Change of \ac{RSSI} with the blockage and proactive handover in Scenario 6.}
		\label{fig: scenario6}
	\end{subfigure}\hfill
	\caption{\ac{RSSI} of the proposed framework and baseline with the blockage and proactive handover.}
	\label{fig: RSSI_change}
\end{figure*}

In Fig. \ref{fig: RSSI_change}, we show the performance of our model with $\flopbound' = 40\%$ and Baseline [MHA] in Scenarios 5 and 6. We emulate the handover by using the measured \ac{RSSI} between the vehicle and \ac{RSU} 2. Blockages happen in the time frame 24 and 21 in Scenarios 5 and 6, respectively. Both models can accurately predict blockages. However, Baseline [MHA] takes more than 300 ms to make a prediction, thereby making an obsolete decision. Hence, it performs handover after the blockage has happened. Meanwhile, our model can finish the inference 83\% faster than Baseline [MHA] by processing only important tokens. Hence, our model does not show severe \ac{RSSI} damage like Baseline [MHA]. 

\section{Conclusion} \label{sec: tokyo_conclusion}

In this paper, we have developed an efficient and fast multi-modal transformer inference framework that processes only important tokens for wireless communication tasks. We have formulated an optimization problem to minimize the loss function by controlling the number of tokens to be processed  under a target computational budget. To solve this problem, we first have projected all modality data into the same embedding dimension. Then, we have utilized token-level routing to learn the importance of each token in the same embedding dimension and process only important tokens. To optimize the number of processed tokens, we have defined a learnable keep ratio for each layer. Simulation results have validated that our model can improve the inference latency significantly with minimal performance loss on two mmWave beamforming tasks. We have also shown that our model can be practically deployed for proactive handover on our real-world multi-modal mmWave handover datasets. In essence, this work presents the first multi-modal transformer inference framework that can balance the tradeoff between performance and latency for practical wireless communication tasks. For future work, we plan to improve robustness of the proposed framework under distribution shifts, including deployment in unseen wireless environments, different carrier frequencies, and unseen blockage types.
	
\bibliographystyle{IEEEtran}
\bibliography{Bibtex/StringDefinitions,Bibtex/IEEEabrv,Bibtex/mybib}

\begin{appendix}
\subsection{Inference latency breakdown}
We provide a detailed breakdown of total inference latency in Tables \ref{tab: latency_breakdown_1} and \ref{tab: latency_breakdown_2} with $\flopbound'=50\%$. We can observe that the overhead from the routers is negligible compared to the complexity of the encoder blocks because the routers simply have two linear layers and GELU activation. We can also see that the tokenizers also have negligible overhead compared to that of encoder blocks. Hence, \ac{FLOPs} can be an appropriate proxy for our framework because encoder blocks have the most dominant \ac{FLOPs}. 
	\begin{table*}[t!]
	\centering
	\begin{tabular}{|c|c|c|c|}
		\hline
		\textbf{} & \textbf{Token routers} & \textbf{Tokenizers} & \textbf{Encoder blocks}  \\ \hline
		Latency [ms]   & $3.04$          & $1.26$         & 42.6       \\ \hline
		
	\end{tabular}
	\caption{Breakdown of the total inference latency on \cite{alkhateeb2023deepsense} and \cite{park2025resource} with $\flopbound'=50\%$.}
	\label{tab: latency_breakdown_1}
\end{table*}
\begin{table*}[t!]
	\centering
	\begin{tabular}{|c|c|c|c|}
		\hline
		\textbf{} & \textbf{Token routers} & \textbf{Tokenizers} & \textbf{Encoder blocks}  \\ \hline
		Latency [ms]   & $6.4$          & $5.5$         & 89.93      \\ \hline
		
	\end{tabular}
	\caption{Breakdown of the total inference latency on our collected data with $\flopbound'=50\%$.}
	\label{tab: latency_breakdown_2}
\end{table*}

\begin{table*}[t!] \smaller
	\centering
	\begin{tabular}{|c|c|c|c|c|c|c|}
		\hline
		\textbf{Methods} &
		\textbf{Accuracy {[}\%{]}} &
		\textbf{Latency {[}ms{]}} &
		\textbf{Memory {[}GB{]}} &
		\textbf{FLOPs {[}G{]}} \\ \hline
		[MHA] & $94.30\pm4.7$          & 335.93         & 1.292          &151.12 \\ \hline
		[MHA] + FP16 & $93.97\pm4.5$ & 145.26 & 0.569 &  151.12 \\ \hline
		[MHA] + Pruning & $60.15\pm7.5$ & 335.90 & 1.292 &  80.63 \\ \hline
		($\flopbound' = 40\%$) &  $93.69\pm5.1$         & 77.07 & 1.074 & 29.67  \\ 	 \hline
		($\flopbound' = 40\%$) + FP16 & $93.34\pm5.0$ & \textbf{39.32}  & 0.507 & 29.67  \\ 	 \hline
		($\flopbound' = 40\%$) + Pruning & $78.49\pm8.8$  & 77.05  & 1.074 & 15.83  \\ 	 \hline
	\end{tabular}
	\caption{Impact of FP16 quantization and structured pruning at sparsity of 50\% on our collected dataset.} 
	\label{tab: tokyo_quantprune}
\end{table*}

\begin{table*}[t!] \smaller
	\centering
	\begin{tabular}{|c|c|c|c|c|c|c|}
		\hline
		\textbf{Methods} &
		\textbf{Top-1 {[}\%{]}} &
		\textbf{Top-3 {[}\%{]}} &
		\textbf{Top-5 {[}\%{]}} &
		\textbf{Latency {[}ms{]}} &
		\textbf{Memory {[}GB{]}} &
		\textbf{FLOPs {[}G{]}} \\ \hline
		[MHA] & $47.5\pm0.9$  & $80.9\pm1.0$  & $91.73\pm0.5 $& 158.14 & 0.797 &  292.19 \\ \hline
		[MHA] + BF16 & $47.28\pm1.0$  & $80.1\pm1.0$  & $91.65\pm0.9 $& 48.23 & 0.452 &  292.19 \\ \hline
		[MHA] + Pruning & $9.15\pm1.3$  & $24.55\pm3.1$  & $32.95\pm3.21 $& 158.10 & 0.797 &  155.53 \\ \hline
		($\flopbound' = 50\%$) &  $47.84\pm0.8$ & $81.8\pm0.9$  & $92.08\pm0.5$ & 46.81  & 0.592 & 75.65  \\ 	 \hline
		($\flopbound' = 50\%$) + FP16 & $47.54\pm0.9$ & $81.42\pm0.8$  & $91.92\pm0.6$ & \textbf{16.82}  & 0.372 & 75.65  \\ 	 \hline
		($\flopbound' = 50\%$) + Pruning & $19.96\pm3.8$  & $37.67\pm8.3$  & $43.95\pm8.3$ & 46.81  & 0.592 & 40.21  \\ 	 \hline
	\end{tabular}
	\caption{Impact of BF16 quantization and structured pruning at sparsity of 50\% on \cite{alkhateeb2023deepsense}.} 
	\label{tab: quantprune}
\end{table*}

\subsection{Complexity of the keep-ratio optimization}
\begin{figure}[t] 		
	\centering 				%
	\includegraphics[width=0.99\linewidth]{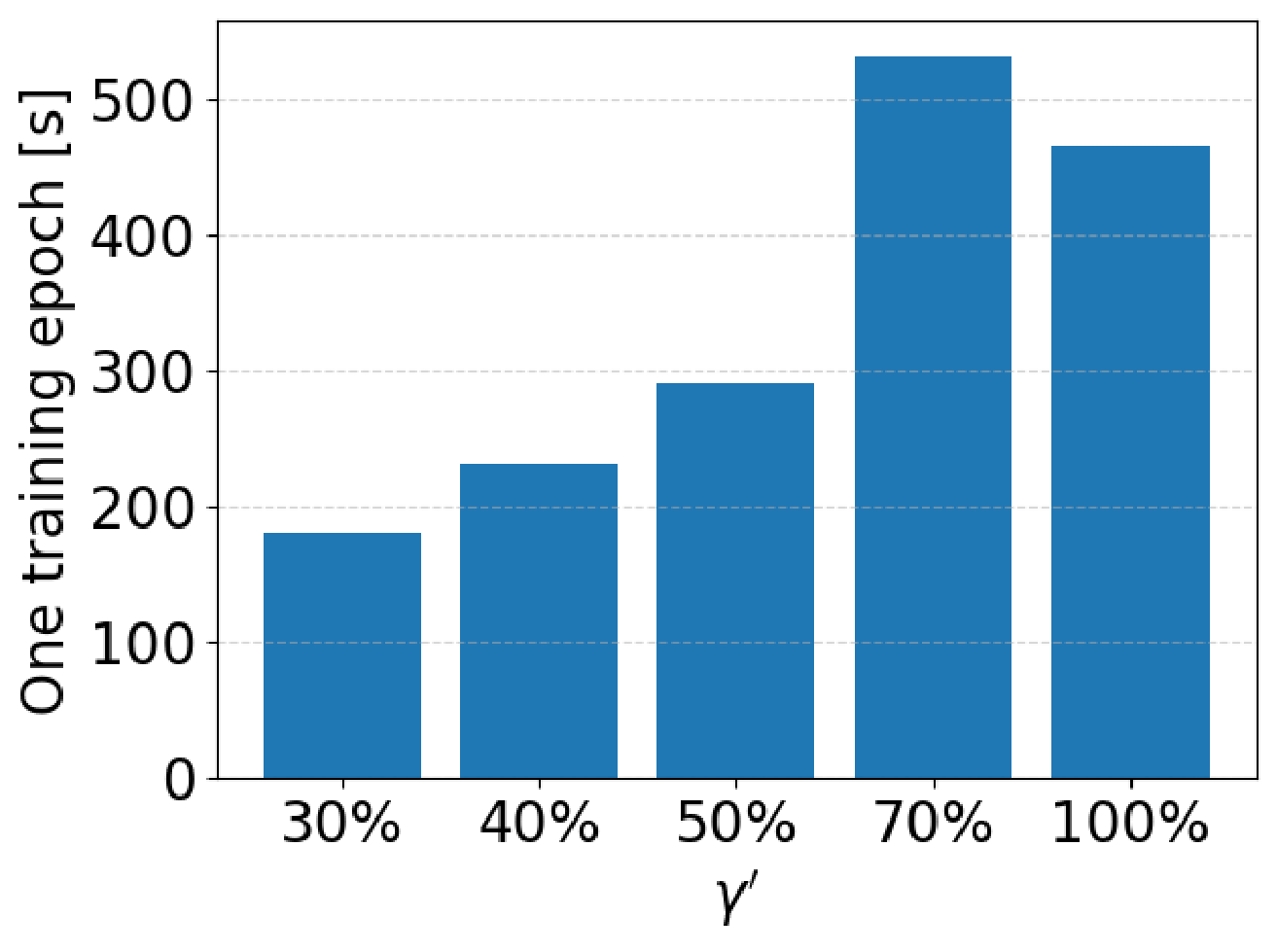}
	\caption{Time for one training epoch with varying $\flopbound'$ on the DeepSense 6G dataset \cite{alkhateeb2023deepsense}.}
	\label{fig: training_epoch}
\end{figure}

Consider the Baseline [MHA] whose computational complexity can be given by $\mathcal{O}(\tokentotalnum^2)$, where $\tokentotalnum$ is the total number of tokens. Essentially, the computational complexity of the keep-ratio optimization can be approximated as $\mathcal{O}(2\topk^2)$. Hence, we still improve the computational complexity of the training when $\topk < \frac{\tokentotalnum}{\sqrt{2}}$. We visualize this observation in Fig. \ref{fig: training_epoch} with one H100 GPU on the DeepSense 6G dataset \cite{alkhateeb2023deepsense}. From Fig. \ref{fig: training_epoch}, we can observe that $\flopbound' = 50\%$ reduces the time for one epoch by 37.6\% compared $\flopbound' = 100\%$ \cite{cui2024sensing}. Meanwhile, we can see that $\flopbound' = 70\%$ slows one training epoch by $15\%$ compared to $\flopbound' = 100\%$. Hence, the keep-ratio optimization can reduce the training time for multi-modal transformers with small $\flopbound'$. 

\subsection{Quantization and Pruning}
In this subsection, we investigate the impact of quantization and pruning on our approach. Note that quantization and pruning are orthogonal to our approach because we are optimizing the number of tokens to be processed. Meanwhile, quantization and pruning consider model weights and activations without controlling the number of tokens. Hence, we can further improve our approach by adding quantization or pruning. We consider lightweight quantization and pruning methods. Specifically, we consider zero-shot post-training quantization and pruning provided by Pytorch. For the quantization, we quantize our trained models into BF16 or FP16 format based on GPUs \cite{quantizaton}. For pruning, we perform magnitude-based structured $l_2$ pruning \cite{pruning} by setting target sparsity as $50\%$. In Table \ref{tab: quantprune}, we show the impact of quantization and pruning on Baseline [MHA] and our proposed approach with $\flopbound'=50\%$ on the DeepSense 6G dataset \cite{alkhateeb2023deepsense} with an NVIDIA A30 GPU.  

	From Table \ref{tab: quantprune}, we can see that quantization with BF16 can significantly reduce latency and memory usage with negligible performance loss. Meanwhile, pruning does not increase the resource-efficiency and degrade the performance. Pruning generally requires a special kernel that supports sparsity patterns to actually reduce inference latency. Moreover, the pruning method uses a mask to make pruned parameters as zero, and save them as FP32, thereby yielding zero memory improvement. Moreover, most CUDA kernels still compute all multiplications with zeros. Hence, pruning does not reduce the inference while theoretical \ac{FLOPs} can be reduced due to masked parameters.

	In Table \ref{tab: tokyo_quantprune}, we provide the impact of the quantization and pruning on our collected dataset on NVIDIA T4 GPU. We used FP16 for quantizaton because T4 GPUs does not support BF16 format. From Table \ref{tab: tokyo_quantprune}, we can see that the quantization can significantly reduce the latency and memory usage with negligible performance loss. Meanwhile, the pruning does not provide meaningful improvements. We can also observe that the improvement of the inference latency with the quantization is smaller than that of Table \ref{tab: quantprune}. This is because A30 GPUs have more tensor cores and higher computing capabilities. In contrast, T4 GPUs have smaller tensor cores and lower memory bandwidth.

\end{appendix}

\end{document}

%% file: Support/acronym.tex
\acrodef{IoT}{Internet of Things}
\acrodef{BS}{base station}
\acrodef{RSS}{received signal strength}
\acrodef{MSE}{mean square error}
\acrodef{MAE}{mean absolute error}
\acrodef{RMSE}{root mean square error}
\acrodef{LoS}{line-of-sight}
\acrodef{NLoS}{non-line-of-sight}
\acrodef{CNN}{convolutional neural network}
\acrodef{MLP}{Multilayer perceptron}
\acrodef{ML}{machine learning}
\acrodef{FLOPs}{floating point operations}
\acrodef{MoD}{mixture-of-depth}
\acrodef{RSSI}{received signal strength index}
\acrodef{RSU}{road side unit}
\acrodef{mmWave}{millimeter wave}
\acrodef{V2I}{vehicle-to-infrastructure}
\acrodef{NLP}{natural language processing}

%% file: Support/defmetric_thesis.tex
\usepackage{color}
\usepackage{dsfont}
\usepackage{bbm}

\newtheorem{theorem}{Theorem}
\newtheorem{lemma}{Lemma}

\newcommand{\E}{\mathbb{E}}
\renewcommand{\P}{\mathbb{P}}
\newcommand{\ka}{\nonumber \\}

\newcommand{\tokentotalnum}{N}
\newcommand{\topk}{K}
\newcommand{\tokyomodalitynum}{M}
\newcommand{\tokyodimension}{d}
\newcommand{\tokyolayernum}{L}

\newcommand{\multimodalin}{\boldsymbol{X}}
\newcommand{\rawmodalin}{\boldsymbol{I}}
\newcommand{\out}{\boldsymbol{Y}}
\newcommand{\nn}{\boldsymbol{\Theta}}
\newcommand{\tokyoloss}{\mathcal{L}}
\newcommand{\flopbound}{\gamma}
\newcommand{\score}{\boldsymbol{s}}
\newcommand{\transformer}{E}
\newcommand{\keepratio}{r}
\newcommand{\tokyonumatennta}{Q}
\newcommand{\tokyoframe}{\tau}
\newcommand{\codebook}{\mathcal{F}}
\newcommand{\beam}{\boldsymbol{f}}
\newcommand{\symboldim}{M}
\newcommand{\tokyorss}{S}
\newcommand{\tokyotxpower}{R}
\newcommand{\tokyopower}{P}
\newcommand{\tokyolink}{\epsilon}
\newcommand{\tokyodistance}{l}

%% file: Bibtex/IEEEabrv.bib
@STRING{IEEE_J_VT         = "{IEEE} Trans. Veh. Technol."}

@STRING{IEEE_J_STSP       = "{IEEE} J. Sel. Topics Signal Process."}

@STRING{IEEE_J_WCOML       = "{IEEE} Wireless Commun. Lett."}

@STRING{IEEE_J_JSAC       = "{IEEE} J. Sel. Areas Commun."}

@STRING{IEEE_J_WCOM       = "{IEEE} Trans. Wireless Commun."}


%% file: Bibtex/StringDefinitions.bib
@STRING{IEEE_J_WCOML = {IEEE Wireless Commun. Lett.}}

@STRING{AI = {Artif. Intell.}}

@STRING{ICC = {Proc. IEEE Int. Conf. Commun.}}

@STRING{IEEE = {The Institute of Electrical and Electronics Engineers}}

@STRING{VTC = {Proc. IEEE Veh. Technol. Conf.}}


%% file: Bibtex/mybib.bib
@ARTICLE{lidar,
author = {Robosense},
title = {RS-LiDAR User Manual},
journal = {https://github.com/RoboSense-LiDAR/rslidar-sdk}
}

@inproceedings{carlak2024state,
	title={State Consistent Edge-enhanced Perception for Connected and Automated Vehicles},
	author={Carlak, Can and Yu, Bo and Bai, Fan and Mao, Z Morley},
	booktitle={IEEEVehicular Technology Conference (VTC)},
	pages={1--7},
	year={2024},
	month = Oct,
	address= {Washington, DC, USA}
}

@inproceedings{yang2024navigator,
	title={Navigator: A Decentralized Scheduler for Latency-Sensitive AI Workflows},
	author={Yang, Yuting and Merlina, Andrea and Song, Weijia and Yuan, Tiancheng and Birman, Ken and Vitenberg, Roman},
	booktitle={2024 IEEE International Conference on Edge Computing and Communications (EDGE)},
	pages={35--47},
	year={2024},
	month = Jul
}

@inproceedings{zhang2021real,
	title={Real-time end-to-end federated learning: An automotive case study},
	author={Zhang, Hongyi and Bosch, Jan and Olsson, Helena Holmstr{\"o}m},
	booktitle={2021 IEEE 45th Annual Computers, Software, and Applications Conference (COMPSAC)},
	pages={459--468},
	year={2021},
	month = Jul
}

@ARTICLE{pruning,
author = {Pytorch},
title ={},
journal = {docs.pytorch.org/docs/stable/generated/torch.nn.utils.prune.ln-structured}
}

@ARTICLE{quantizaton,
	author = {Pytorch},
	title ={},
journal = {docs.pytorch.org/docs/stable/generated/torch.Tensor.bfloat16}
}

@article{lei2019stochastic,
	title={Stochastic gradient descent for nonconvex learning without bounded gradient assumptions},
	author={Lei, Yunwen and Hu, Ting and Li, Guiying and Tang, Ke},
	journal={IEEE transactions on neural networks and learning systems},
	volume={31},
	number={10},
	pages={4394--4400},
	year={2020},
	month =Oct
}

@article{shang2025multi,
	title={Multi-Modal Beamforming with Model Compression and Modality Generation for V2X Networks},
	author={Shang, Chen and Hoang, Dinh Thai and Yu, Jiadong},
	journal={arXiv preprint arXiv:2506.22469},
	year={2025}
}

@article{tian2023multimodal,
	title={Multimodal transformers for wireless communications: A case study in beam prediction},
	author={Tian, Yu and Zhao, Qiyang and Boukhalfa, Fouzi and Wu, Kebin and Bader, Faouzi and others},
	journal={arXiv preprint arXiv:2309.11811},
	year={2023}
}

@inproceedings{bergner2025token,
	title={Token cropr: Faster vits for quite a few tasks},
	author={Bergner, Benjamin and Lippert, Christoph and Mahendran, Aravindh},
	booktitle={Proceedings of the Computer Vision and Pattern Recognition Conference},
	pages={9740--9750},
	year={2025}
}

@inproceedings{yinunderstanding,
	title={Understanding Straight-Through Estimator in Training Activation Quantized Neural Nets},
	author={Yin, Penghang and Lyu, Jiancheng and Zhang, Shuai and Osher, Stanley and Qi, Yingyong and Xin, Jack},
	booktitle={International Conference on Learning Representations},
	year = {2019}
}

@article{team2024chameleon,
	title={Chameleon: Mixed-modal early-fusion foundation models},
	author={Team, Chameleon},
	journal={arXiv preprint arXiv:2405.09818},
	year={2024}
}

@inproceedings{jang2017categorical,
	title={Categorical Reparameterization with Gumbel-Softmax},
	author={Jang, Eric and Gu, Shixiang and Poole, Ben},
	booktitle={International Conference on Learning Representations},
	year={2017}
}

@ARTICLE{10171192,
	author={Kim, Minsu and Saad, Walid and Mozaffari, Mohammad and Debbah, Mérouane},
	journal={IEEE Transactions on Wireless Communications}, 
	title={Green, Quantized Federated Learning Over Wireless Networks: An Energy-Efficient Design}, 
	year={2024},
	volume={23},
	number={2},
	pages={1386-1402},
	month = Feb}

@INPROCEEDINGS{9838838,
	author={Zhang, Yixin and Cheng, Wenchi and Zhang, Wei},
	booktitle= ICC, 
	title={Dumb {RIS}-Assisted Random Beamforming for Energy Efficiency Enhancement of Wireless Communications}, 
	year={2022},
	month = May,
	address = {Seoul, South Korea},
	pages={129-134},
}

@article{va2016impact,
	title={The impact of beamwidth on temporal channel variation in vehicular channels and its implications},
	author={Va, Vutha and Choi, Junil and Heath, Robert W},
	journal=IEEE_J_VT,
	volume={66},
	number={6},
	pages={5014--5029},
	year={2016},
	month = Jun
}

@ARTICLE{10177877,
	author={Zhang, Yixin and Cheng, Wenchi and Zhang, Wei},
	journal=IEEE_J_WCOM, 
	title={Multiple Access Integrated Adaptive Finite Blocklength for Ultra-Low Delay in {6G} Wireless Networks}, 
	year={2024},
	volume={23},
	number={3},
	pages={1670-1683},
}

@ARTICLE{10929033,
  author={Saad, Walid and Hashash, Omar and Thomas, Christo Kurisummoottil and Chaccour, Christina and Debbah, Mérouane and Mandayam, Narayan and Han, Zhu},
  journal={Proceedings of the IEEE}, 
  title={Artificial General Intelligence ({AGI})-Native Wireless Systems: A Journey Beyond {6G}}, 
  year={2025},
  volume={},
  number={},
  pages={1-39},
}

@article{alkhateeb2023deepsense,
  title={DeepSense {6G}: A large-scale real-world multi-modal sensing and communication dataset},
  author={Alkhateeb, Ahmed and Charan, Gouranga and Osman, Tawfik and Hredzak, Andrew and Morais, Joao and Demirhan, Umut and Srinivas, Nikhil},
  journal={IEEE Communications Magazine},
  volume={61},
  number={9},
  pages={122--128},
  year={2023},
  month = Sep
}

@article{kim2024spafl,
  title={SpaFL: Communication-efficient federated learning with sparse models and low computational overhead},
  author={Kim, Minsu and Saad, Walid and Debbah, Merouane and Hong, Choong S},
  journal={Advances in Neural Information Processing Systems},
  volume={37},
  pages={86500--86527},
  year={2024}
}

@article{yang2023environment,
  title={Environment semantics aided wireless communications: A case study of mmWave beam prediction and blockage prediction},
  author={Yang, Yuwen and Gao, Feifei and Tao, Xiaoming and Liu, Guangyi and Pan, Chengkang},
  journal=IEEE_J_JSAC,
  volume={41},
  number={7},
  pages={2025--2040},
  year={2023},
  month = Jul
}

@ARTICLE{9923616,
	author={Xu, Weihua and Gao, Feifei and Tao, Xiaoming and Zhang, Jianhua and Alkhateeb, Ahmed},
	journal= IEEE_J_WCOM, 
	title={Computer Vision Aided mmWave Beam Alignment in V2X Communications}, 
	year={2023},
	volume={22},
	number={4},
	pages={2699-2714},
	month = Apr}

@ARTICLE{10949588,
	author={Bai, Lu and Huang, Ziwei and Sun, Mingran and Cheng, Xiang and Cui, Lizhen},
	journal={IEEE Communications Surveys and Tutorials}, 
	title={Multi-Modal Intelligent Channel Modeling: A New Modeling Paradigm via Synesthesia of Machines}, 
	year={2025},
	month = Apr}

@inproceedings{bolya2023token,
	title={Token Merging: Your ViT But Faster},
	author={Bolya, Daniel and Fu, Cheng-Yang and Dai, Xiaoliang and Zhang, Peizhao and Feichtenhofer, Christoph and Hoffman, Judy},
	booktitle={International Conference on Learning Representations},
	year={2023},
	month = May,
	address = {Kigali, Rwanda}
}

@article{park2025resource,
	title={Resource-Efficient Beam Prediction in mmWave Communications with Multimodal Realistic Simulation Framework},
	author={Park, Yu Min and Tun, Yan Kyaw and Saad, Walid and Hong, Choong Seon},
	journal={arXiv preprint arXiv:2504.05187},
	year={2025}
}

@article{kwon2022fast,
	title={A fast post-training pruning framework for transformers},
	author={Kwon, Woosuk and Kim, Sehoon and Mahoney, Michael W and Hassoun, Joseph and Keutzer, Kurt and Gholami, Amir},
	journal={Advances in Neural Information Processing Systems},
	volume={35},
	pages={24101--24116},
	year={2022}
}

@article{zheng2022savit,
	title={Savit: Structure-aware vision transformer pruning via collaborative optimization},
	author={Zheng, Chuanyang and Zhang, Kai and Yang, Zhi and Tan, Wenming and Xiao, Jun and Ren, Ye and Pu, Shiliang and others},
	journal={Advances in Neural Information Processing Systems},
	volume={35},
	pages={9010--9023},
	year={2022}
}

@article{park2023accurate,
	title={Accurate retraining-free pruning for pretrained encoder-based language models},
	author={Park, Seungcheol and Choi, Hojun and Kang, U},
	journal={International Conference on Learning Representations},
	year={2024}
}

@inproceedings{meng2024falcon,
	title={FALCON: FLOP-aware combinatorial optimization for neural network pruning},
	author={Meng, Xiang and Chen, Wenyu and Benbaki, Riade and Mazumder, Rahul},
	booktitle={International Conference on Artificial Intelligence and Statistics},
	pages={4384--4392},
	year={2024},
	organization={PMLR}
}

@article{mod,
	title={Mixture-of-depths: Dynamically allocating compute in transformer-based language models},
	author={Raposo, David and Ritter, Sam and Richards, Blake and Lillicrap, Timothy and Humphreys, Peter Conway and Santoro, Adam},
	journal={arXiv preprint arXiv:2404.02258},
	year={2024}
}

@article{wu2024videollm,
	title={Videollm-mod: Efficient video-language streaming with mixture-of-depths vision computation},
	author={Wu, Shiwei and Chen, Joya and Lin, Kevin Qinghong and Wang, Qimeng and Gao, Yan and Xu, Qianli and Xu, Tong and Hu, Yao and Chen, Enhong and Shou, Mike Zheng},
	journal={Advances in Neural Information Processing Systems},
	volume={37},
	pages={109922--109947},
	year={2024}
}

@inproceedings{you2025layer,
	title={Layer-and Timestep-Adaptive Differentiable Token Compression Ratios for Efficient Diffusion Transformers},
	author={You, Haoran and Barnes, Connelly and Zhou, Yuqian and Kang, Yan and Du, Zhenbang and Zhou, Wei and Zhang, Lingzhi and Nitzan, Yotam and Liu, Xiaoyang and Lin, Zhe and others},
	booktitle={Proceedings of the Computer Vision and Pattern Recognition Conference},
	pages={18072--18082},
	year={2025}
}

@article{cui2024sensing,
	title={Sensing-assisted high reliable communication: A transformer-based beamforming approach},
	author={Cui, Yuanhao and Nie, Jiali and Cao, Xiaowen and Yu, Tiankuo and Zou, Jiaqi and Mu, Junsheng and Jing, Xiaojun},
	journal= IEEE_J_STSP,
	volume={18},
	number={5},
	pages={782--795},
	year={2024},
	month = Jul
}

@article{rao2021dynamicvit,
	title={Dynamicvit: Efficient vision transformers with dynamic token sparsification},
	author={Rao, Yongming and Zhao, Wenliang and Liu, Benlin and Lu, Jiwen and Zhou, Jie and Hsieh, Cho-Jui},
	journal={Advances in neural information processing systems},
	volume={34},
	pages={13937--13949},
	year={2021}
}

@article{lin2021memory,
	title={Memory-efficient patch-based inference for tiny deep learning},
	author={Lin, Ji and Chen, Wei-Ming and Cai, Han and Gan, Chuang and Han, Song},
	journal={Advances in Neural Information Processing Systems},
	volume={34},
	pages={2346--2358},
	year={2021}
}

@article{
	bonnaerens2023learned,
	title={Learned Thresholds Token Merging and Pruning for Vision Transformers},
	author={Maxim Bonnaerens and Joni Dambre},
	journal={Transactions on Machine Learning Research},
	year={2023},

}

@ARTICLE{10735366,
	author={Ghassemi, Mohammad and Zhang, Han and Afana, Ali and Sediq, Akram Bin and Erol-Kantarci, Melike},
	journal={IEEE Networking Letters}, 
	title={Multi-Modal Transformer and Reinforcement Learning-Based Beam Management}, 
	year={2024},
	volume={6},
	number={4},
	pages={222-226},
	month = Dec}

@ARTICLE{11018220,
		author={Raha, Avi Deb and Kim, Kitae and Adhikary, Apurba and Gain, Mrityunjoy and Han, Zhu and Hong, Choong Seon},
		journal= IEEE_J_VT, 
		title={Advancing Ultra-Reliable 6 G: Transformer and Semantic Localization Empowered Robust Beamforming in Millimeter-Wave Communications}, 
		year={2025},
		volume={},
		number={},
		pages={1-16}}

@ARTICLE{10660494,
	author={Gharsallah, Ghazi and Kaddoum, Georges},
	journal={IEEE Open Journal of the Communications Society}, 
	title={{MVX-ViT}: Multimodal Collaborative Perception for {6G} {V2X} Network Management Decisions Using Vision Transformer}, 
	year={2024},
	volume={5},
	pages={5619-5634},
	month = Aug}

@ARTICLE{10680020,
	author={Gharsallah, Ghazi and Kaddoum, Georges},
	journal={IEEE Access}, 
	title={{ViT LoS V2X}: Vision Transformers for Environment-Aware {LoS} Blockage Prediction for {6G} Vehicular Networks}, 
	year={2024},
	volume={12},
	pages={133569-133583},
	month = Sep}

@inproceedings{
	gotz2025efficient,
	title={Efficient Time Series Processing for Transformers and State-Space Models through Token Merging},
	author={Leon G{\"o}tz and Marcel Kollovieh and Stephan G{\"u}nnemann and Leo Schwinn},
	booktitle={International Conference on Machine Learning},
	year={2025}
}

@article{khorsandmanesh2024beam,
	title={Beam Coherence Time Analysis for Mobile Wideband mmWave Point-to-Point MIMO Channels},
	author={Khorsandmanesh, Yasaman and Bj{\"o}rnson, Emil and Jald{\'e}n, Joakim and Lindoff, Bengt},
	journal=IEEE_J_WCOML,
	volume={13},
	number={6},
	pages={1546--1550},
	year={2024},
	month = Jun
}
